\documentclass[11pt,a4]{article}
\usepackage[T1]{fontenc}
\usepackage[latin1]{inputenc}
\usepackage{graphicx}
 \usepackage{fancyhdr,lastpage}
 \usepackage{amsmath}
 \usepackage{latexsym}
 \usepackage{url}

\begin{document}

 \renewcommand{\footnoterule}{\vspace{0.5cm}%
 \rule{2.5in}{0.4pt} \vspace{0.3cm}}

\newcounter{myfn}[page]
\renewcommand{\thefootnote}{\fnsymbol{footnote}}
\newcommand{\myfootnote}[1]{
  \setcounter{footnote}{\value{myfn}}%
  \footnote{#1}\stepcounter{myfn}} \newcommand{\fn}[1]{\myfootnote{
    #1}}
 \newcommand{\mean}[1]{\left\langle #1 \right\rangle}
 \newcommand{\abs}[1]{\left| #1 \right|}
 \newcommand{\la}{\langle}
 \newcommand{\ra}{\rangle}
 \newcommand{\RA}{\Rightarrow}
 \newcommand{\tet}{\vartheta}
 \newcommand{\eps}{\varepsilon}
 \newcommand{\bbox}[1]{\mbox{\boldmath $#1$}}
 \newcommand{\ul}[1]{\underline{#1}}
 \newcommand{\ol}[1]{\overline{#1}}
 \newcommand{\non}{\nonumber \\}
 \newcommand{\no}{\nonumber}
 \newcommand{\eqn}[1]{eq. (\ref{#1})}
 \newcommand{\Eqn}[1]{Eq. (\ref{#1})}
 \newcommand{\eqs}[2]{eqs. (\ref{#1}), (\ref{#2})}
 \newcommand{\pics}[2]{Figs. \ref{#1}, \ref{#2}}
 \newcommand{\pic}[1]{Fig. \ref{#1}}
 \newcommand{\sect}[1]{Sect. \ref{#1}}
 \newcommand{\name}[1]{{\rm #1}}

\newcommand{\rb}[1]{\raisebox{-1 ex}{#1}}
\newcommand{\av}[1]{\left< #1 \right>}
\newcommand{\abst}[0]{\rule[-1.5 ex]{0 ex}{4 ex}}
\newcommand{\fref}[1]{Fig.~\ref{#1}}
\newcommand{\tref}[1]{Tab.~\ref{#1}}
\newcommand{\eref}[1]{Eq.~(\ref{#1})}
\newcommand{\sref}[1]{Section~\ref{#1}}
\newcommand{\aref}[1]{Appendix~\ref{#1}}

\newcommand{\vol}[1]{{\bf #1}}
 \newcommand{\et}{{\it et al.}}
 \newcommand{\D}{\displaystyle}
 \newcommand{\T}{\textstyle}
 \newcommand{\SC}{\scriptstyle}
 \newcommand{\SSC}{\scriptscriptstyle}
 \renewcommand{\textfraction}{0.05}
 \renewcommand{\topfraction}{0.95}
 \renewcommand{\bottomfraction}{0.95}
 \renewcommand{\floatpagefraction}{0.95}

 \begin{center}

   \textbf{\Large The network of global corporate control}\\[5mm]

{\large \bf  Stefania Vitali$^{1}$, James B. Glattfelder$^{1}$, and Stefano Battiston$^{1}$$^{\star}$}

\begin{quote}
\begin{itemize}
\item[$^{1}$] \emph{Chair of Systems Design, ETH Zurich, Kreuzplatz 5,
    8032 Zurich, Switzerland}, {$^{\star}$corresponding author, email:
    sbattiston@ethz.ch}
\end{itemize}
\end{quote}

\end{center}

\section*{Abstract}
The structure of the control network of transnational corporations
affects global market competition and financial stability. So far,
only small national samples were studied and there was no appropriate
methodology to assess control globally. We present the first
investigation of the architecture of the international ownership
network, along with the computation of the control held by each global
player. We find that transnational corporations form a giant bow-tie
structure and that a large portion of control flows to a small
tightly-knit core of financial institutions. This core can be seen as
an economic ``super-entity'' that raises new important issues both for
researchers and policy makers.

\section*{Introduction}
A common intuition among scholars and in the media sees the global
economy as being dominated by a handful of powerful transnational
corporations (TNCs). However, this has not been confirmed or rejected
with explicit numbers.  A quantitative investigation is not a trivial
task because firms may exert control over other firms via a web of
direct and indirect ownership relations which extends over many
countries. Therefore, a complex network analysis \cite{Barabasi1999}
is needed in order to uncover the structure of control and its
implications. Recently, economic networks have attracted growing
attention \cite{schweitzerea09}, e.g., networks of trade
\cite{fagiolo.ea09}, products \cite{hidalgo2009building}, credit
\cite{boss2004nti,iori2008nai}, stock prices \cite{PhysRevE.68.046130}
and boards of directors \cite{strogatz2001ecn,Battiston2004spc}. This
literature has also analyzed ownership networks
\cite{kogut.ea01,GlattfelderBattiston2009BackboneComplexNetworks}, but
has neglected the structure of control at a global level. Even the
corporate governance literature has only studied small national
business groups \cite{Granovetter1995crb}. Certainly, it is intuitive
that every large corporation has a pyramid of subsidiaries below and a
number of shareholders above. However, economic theory does not offer
models that predict how TNCs globally connect to each other. Three
alternative hypotheses can be formulated. TNCs may remain isolated,
cluster in separated coalitions, or form a giant connected component,
possibly with a core-periphery structure. So far, this issue has
remained unaddressed, notwithstanding its important implications for
policy making.  Indeed, mutual ownership relations among firms within
the same sector can, in some cases, jeopardize market competition
\cite{obrien.ea99,gilo.ea06}. Moreover, linkages among financial
institutions have been recognized to have ambiguous effects on their
financial fragility
\cite{allen2000financial,stiglitz2010risk}. Verifying to what extent
these implications hold true in the global economy is \textit{per se}
an unexplored field of research and is beyond the scope of this
article. However, a necessary precondition to such investigations is
to uncover the worldwide structure of corporate control. This was
never performed before and it is the aim of the present work.

\begin{figure}[t]
  \begin{center}

  \begin{minipage}{0.4\textwidth}
    \flushleft
    \textsf{\textbf{A}}\\
    \centering
    \includegraphics[width=0.725\textwidth]{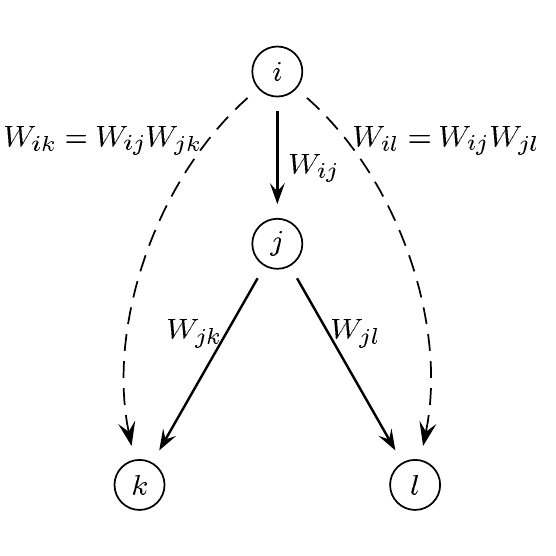}
  \end{minipage}
  \begin{minipage}{0.4\textwidth}
    \flushleft 
    \textsf{\textbf{B}} \\
    \centering
    \includegraphics[width=0.725\textwidth]{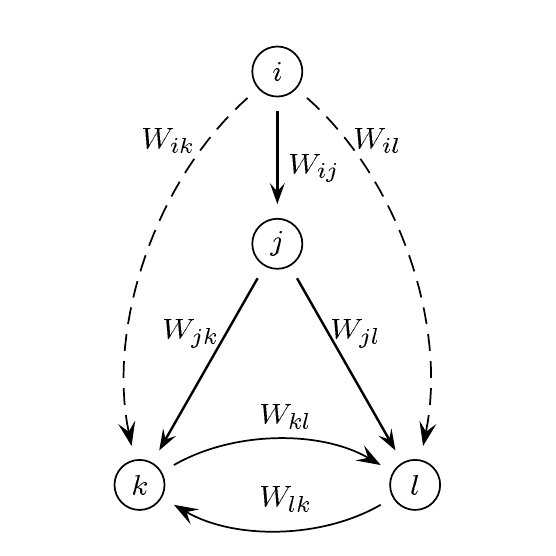}
  \end{minipage}
  \begin{minipage}{0.4\textwidth}
    \flushleft
    \textsf{\textbf{C}}\\
\centering
    \includegraphics[width=0.725\textwidth]{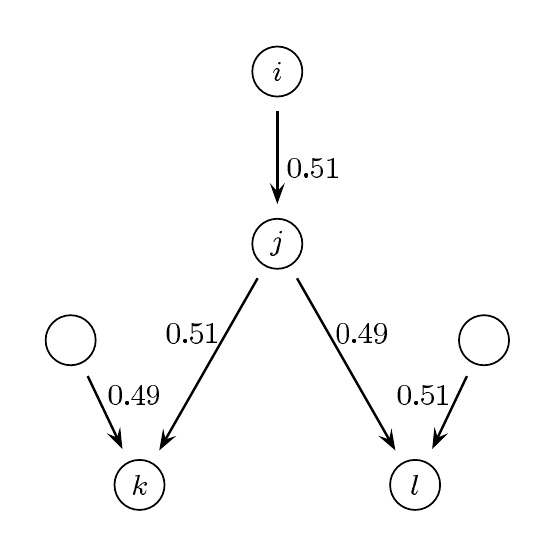}
  \end{minipage}
  \begin{minipage}{0.4\textwidth}
    \flushleft
    \textsf{\textbf{D}}\\
    \centering 
    \includegraphics[width=0.725\textwidth]{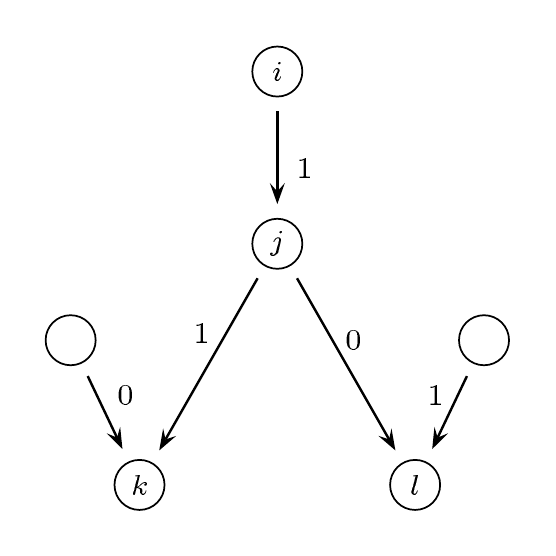}
  \end{minipage} 
  \caption{{\bf Ownership and Control.} ({\bf{\textsf{A\&B}}}) Direct
    and indirect ownership. ({\textsf{A}}) Firm $i$ has $W_{ij}$
    percent of direct ownership in firm $j$. Through $j$, it has also
    an indirect ownership in $k$ and $l$.  ({\textsf{B}}) With cycles
    one has to take into account the recursive paths, see SI Appendix,
    Sec. 3.1. ({\bf{\textsf{C\&D}}}) Threshold model. ({\textsf{C}})
    Percentages of ownership are indicated along the
    links. ({\textsf{D}}) If a shareholder has ownership exceeding a
    threshold (e.g. $50\%$), it has full control (100\%) and the
    others have none (0\%). More conservative model of control are
    also considered see SI Appendix, Sec. 3.1.  }
  \label{fig:ownershipcontrol}
\end{center}
\end{figure}

\section*{Methods}
Ownership refers to a person or a firm owning another firm entirely or
partially. Let $W$ denote the ownership matrix, where the component
$W_{ij}\in [0,\,1]$ is the percentage of ownership that the owner (or
\textit{shareholder}) $i$ holds in firm $j$. This corresponds to a
directed weighted graph with firms represented as nodes and ownership
ties as links. If, in turn, firm $j$ owns $W_{jl}$ shares of firm $l$,
then firm $i$ has an \textit{indirect ownership} of firm $l$ (Fig.\
\ref{fig:ownershipcontrol} A). In the simplest case, this amounts
trivially to the product of the shares of direct ownership
$W_{ij}W_{jl}$. If we now consider the economic value $v$ of firms
(e.g., operating revenue in USD), an amount $W_{ij}v_j$ is associated
to $i$ in the direct case, and $W_{ij}W_{jl}v_l$ in the indirect
case. This computation can be extended to a generic graph, with some
important caveats \cite[SI Appendix, Secs. 3.1 and 3.2]{brioschi.ea89}.

Each shareholder has the right to a fraction of the firm revenue
(dividend) and to a voice in the decision making process (e.g., voting
rights at the shareholder meetings). Thus the larger the ownership
share $W_{ij}$ in a firm, the larger is the associated
\textit{control} over it, denoted as $C_{ij}$. Intuitively, control
corresponds to the chances of seeing one's own interest prevailing in
the business strategy of the firm. Control $C_{ij}$ is usually
computed from ownership $W_{ij}$ with a simple threshold rule: the
majority shareholder has full control. In the example of Fig.\
\ref{fig:ownershipcontrol} C, D, this yields $C_{ij}v_j= 1 \, v_j$ in
the direct case and $C_{ij}C_{jl}v_l = 0$ in the indirect case. As a
robustness check, we tested also more conservative models where
minorities keep some control (see SI Appendix, Sec. 3.1). In analogy
to ownership, the extension to a generic graph is the notion of
\textit{network control}: $c^{\textrm{net}}_i = \sum_j C_{ij}v_j +
\sum_j C_{ij}c^{\textrm{net}}_j$. This sums up the value controlled by
$i$ through its shares in $j$, plus the value controlled indirectly
via the network control of $j$. Thus, network control has the meaning
of the total amount of economic value over which $i$ has an influence
(e.g. $c_i^{\textrm{net}}=v_j+v_k$ in Fig.\ \ref{fig:ownershipcontrol}
D).

Because of indirect links, control flows upstream from many firms and
can result in some shareholders becoming very powerful. However,
especially in graphs with many cycles (see Figs. 1 B and S4), the
computation of $c^{\textrm{net}}$, in the basic formulation detailed
above, severely overestimates the control assigned to actors in two
cases: firms that are part of cycles (or cross-shareholding
structures), and shareholders that are upstream of these
structures. An illustration of the problem on a simple network
example, together with the details of the method are provided in SI
Appendix, Secs. 3.2 -- 3.4. A partial solution for small networks was
provided in
\cite{baldone.ea98}. Previous work on large control networks used a
different network construction method and neglected this issue
entirely \cite[SI Appendix, Secs. 2 and
3.5]{GlattfelderBattiston2009BackboneComplexNetworks}. In this paper,
by building on \cite{GlattfelderBattiston2009BackboneComplexNetworks},
we develop a new methodology to overcome the problem of control
overestimation, which can be employed to compute control in large
networks.

\section*{Results}
We start from a list of 43060 TNCs identified according to the OECD
definition, taken from a sample of about 30 million economic actors
contained in the Orbis 2007 database (see SI Appendix, Sec. 2). We
then apply a recursive search (Fig.\ S1 and SI Appendix, Sec. 2) which
singles out, for the first time to our knowledge, the network of all
the ownership pathways originating from and pointing to TNCs (Fig.
S2). The resulting TNC network includes 600508 nodes and 1006987
ownership ties.

\begin{figure}[ht!]
  \begin{center}
    \begin{minipage}{0.5\textwidth}
      \flushleft
      \textsf{\textbf{A}}\\
      \vspace{0.3in} \hspace{0.1in}
      \includegraphics[width=0.7\textwidth]{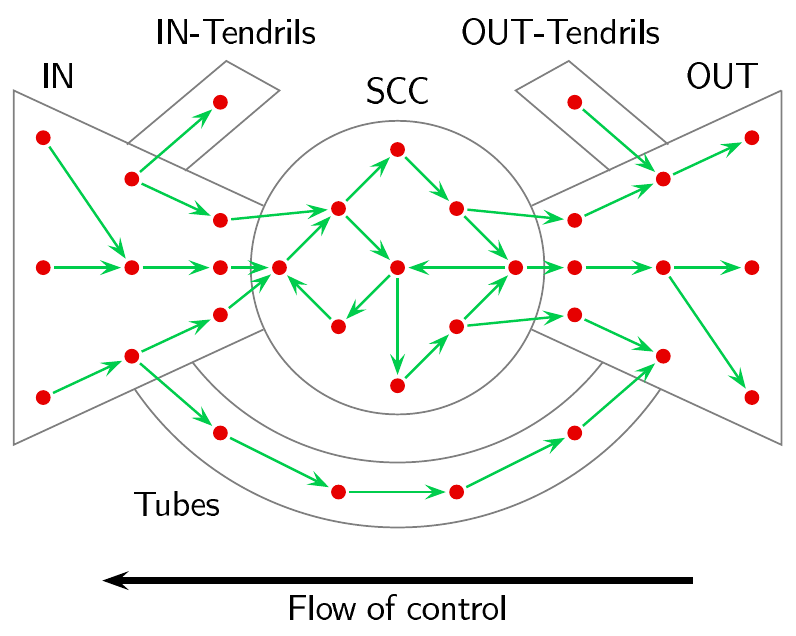}
      \vspace{0.3in}
    \end{minipage}
    \hspace{1em}
    \begin{minipage}{0.45\textwidth}
      \flushleft \hspace{-0.45in}
      \textsf{\textbf{B}} \hspace{11in}\\
      \vspace{1em}
      \includegraphics[width=0.81\textwidth]{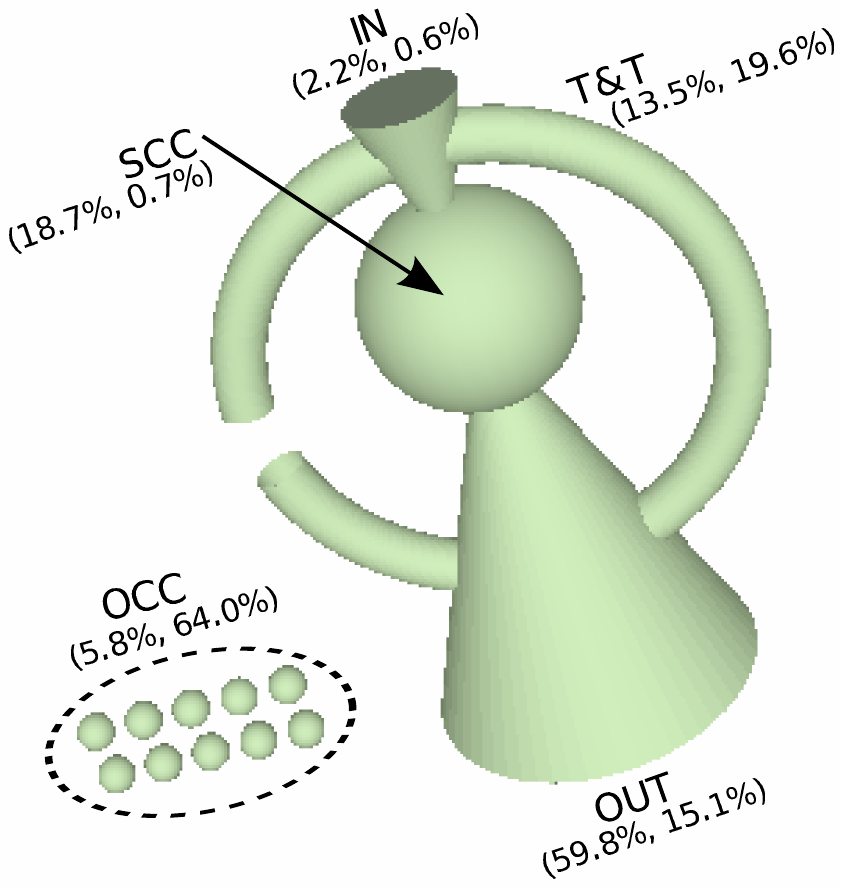}
    \end{minipage}
    \vspace{1em}
    \begin{minipage}{0.5\textwidth}
      \flushleft
      \textsf{\textbf{C}}\\
      \includegraphics[width=0.81\textwidth]{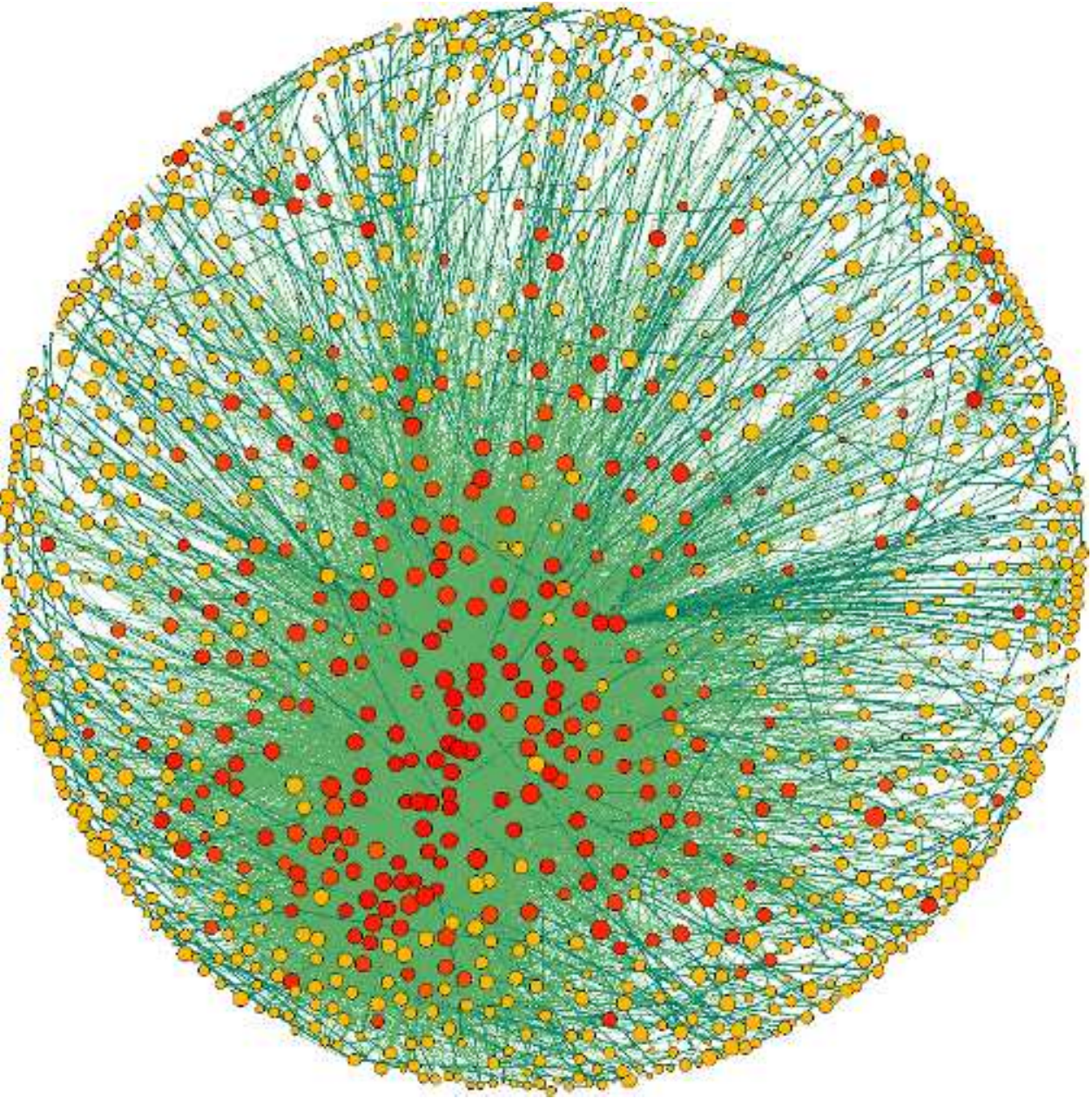}
  \end{minipage}
  \hspace{-2.5em}
  \begin{minipage}{0.5\textwidth}
    \flushleft
    \textsf{\textbf{D}}\\
    \vspace{0.5in}
    \includegraphics[width=1.1\textwidth]{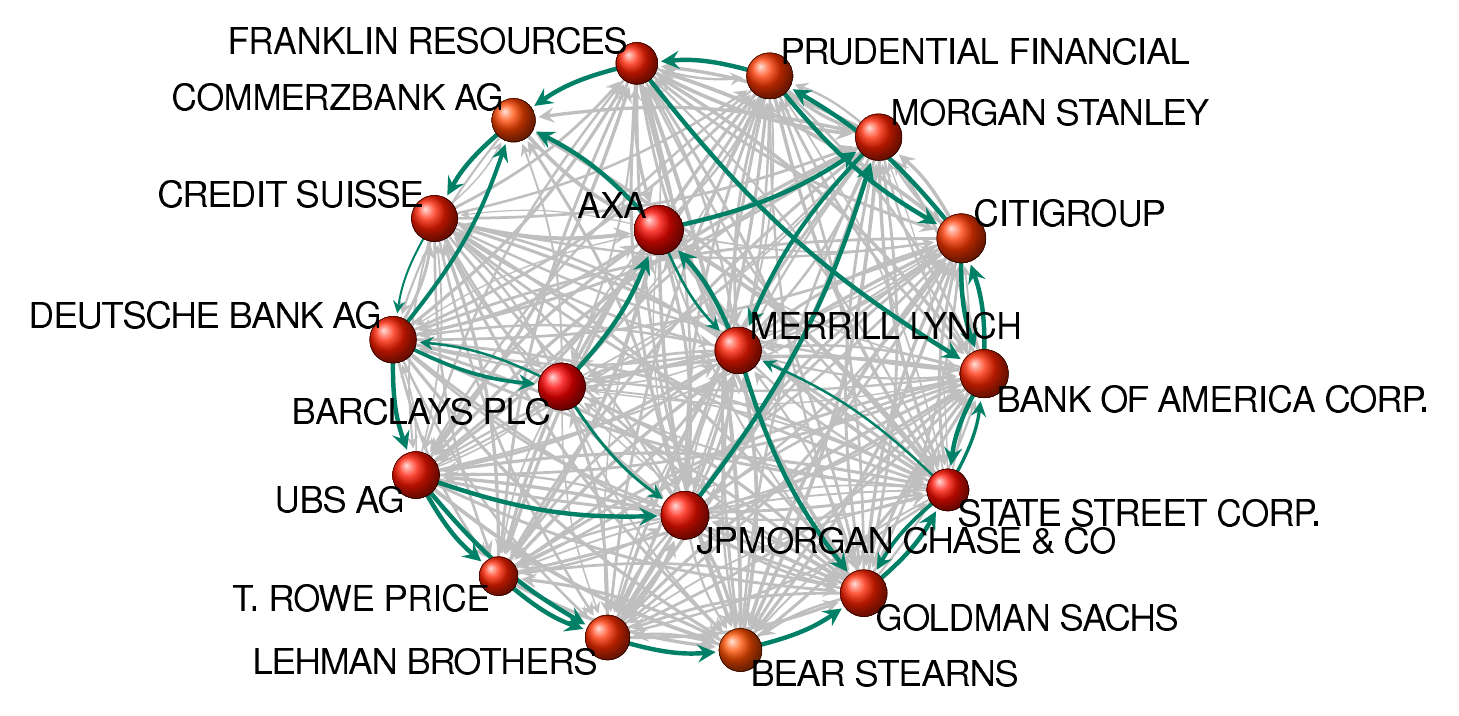}
    \vspace{0.15in}
  \end{minipage}
  \caption{ {\bf Network topology.} ({\bf{\textsf A}}) A bow-tie
    consists of in-section (IN), out-section (OUT), strongly connected
    component or core (SCC), and tubes and tendrils (T\&T).
    ({\bf{\textsf B}}) Bow-tie structure of the largest connected
    component (LCC) and other connected components (OCC). Each section
    volume scales logarithmically with the share of its TNCs operating
    revenue. In parenthesis, percentage of operating revenue and
    number of TNCs, cfr. Table 1. ({\bf{\textsf C}}) SCC layout of the
    SCC (1318 nodes and 12191 links). Node size scales logarithmically
    with operation revenue, node color with network control (from
    yellow to red). Link color scales with weight. ({\bf{\textsf D}})
    Zoom on some major TNCs in the financial sector. Some cycles are
    highlighted.}
  \label{fig:top}
\end{center}
\end{figure}

Notice that this data set fundamentally differs from the ones analysed
in \cite{GlattfelderBattiston2009BackboneComplexNetworks} (which
considered only listed companies in separate countries and their
direct shareholders). Here we are interested in the true global
ownership network and many TNCs are not listed companies (see also SI
Appendix, Sec. 2).

\subsection*{Network Topology}

The computation of control requires a prior analysis of the topology.
In terms of connectivity, the network consists of many small connected
components, but the largest one (3/4 of all nodes) contains all the
top TNCs by economic value, accounting for 94.2\% of the total TNC
operating revenue (Tbl.\ \ref{tab:statsbowtie}). Besides the usual
network statistics (Figs.\ S5, S6), two topological properties are the
most relevant to the focus of this work. The first is the abundance of
cycles of length two (mutual cross-shareholdings) or greater (Fig.  S7
and SI Appendix, Sec. 7), which are well studied motifs in corporate
governance \cite{dietzenbacher.ea08}.  A generalization is a
\textit{strongly connected component} (SCC), i.e., a set of firms in
which every member owns directly and/or indirectly shares in every
other member.  This kind of structures, so far observed only in small
samples, has explanations such as anti-takeover strategies, reduction
of transaction costs, risk sharing, increasing trust and groups of
interest \cite{williamson75}. No matter its origin, however, it
weakens market competition \cite{obrien.ea99, gilo.ea06}. The second
characteristics is that the largest connect component contains only
one dominant strongly connected component (1347 nodes). Thus, similar
to the WWW, the TNC network has a \textit{bow-tie} structure
\cite{broder.ea00} (see Fig.\ \ref{fig:top} A and SI Appendix, Sec. 6).  
Its peculiarity is that the strongly connected component, or
\textit{core}, is very small compared to the other sections of the
bow-tie, and that the out-section is significantly larger than the
in-section and the tubes and tendrils (Fig.\ \ref{fig:top} B and Tbl.\
\ref{tab:statsbowtie}).  The core is also very densely connected, with
members having, on average, ties to 20 other members (Fig.\
\ref{fig:top} C, D). As a result, about 3/4 of the ownership of firms
in the core remains in the hands of firms of the core itself. In other
words, this is a tightly-knit group of corporations that cumulatively
hold the majority share of each other.

\begin{table}[t]
\centering
  \caption{{\bf Bow-tie statistics.} Percentage of total TNC operating revenue
    (OR) and number (\#) of nodes in the sections of the bow-tie
    (acronyms are in Fig. \ref{fig:top}). Economic actors types are:
    shareholders (SH), participated companies (PC).}
  \label{tab:statsbowtie}
  \begin{tabular}{lrrrr} {\em } & { TNC (\#)} & { SH (\#)} &
    { PC (\#)} & { OR (\%)} \\
    \hline
    LCC & 15491 & 47819 & 399696 & 94.17 \\
    IN & 282 & 5205 & 129 & 2.18 \\
    SCC & 295 & 0 & 1023 & 18.68 \\
    OUT & 6488 & 0 & 318073 & 59.85 \\
    T\&T & 8426 & 42614 & 80471 & 13.46 \\
    OCC & 27569 & 29637 & 80296 & 5.83 \\
    \hline 
  \end{tabular} 
\end{table}

Notice that the cross-country analysis of
\cite{GlattfelderBattiston2009BackboneComplexNetworks} found that only
a few of the national ownership networks are bow-ties, and,
importantly, for the Anglo-Saxon countries, the main strongly
connected components are big compared to the network size.

\subsection*{Concentration of Control}

The topological analysis carried out so far does not consider the
diverse economic value of firms.  We thus compute the network control
that economic actors (including TNCs) gain over the TNCs' value
(operating revenue) and we address the question of how much this
control is concentrated and who are the top control holders. See
Fig. S3 for the distribution of control and operating revenue.

\begin{figure}[t]
\begin{center}
  \includegraphics[width=0.56\textwidth,angle=0]{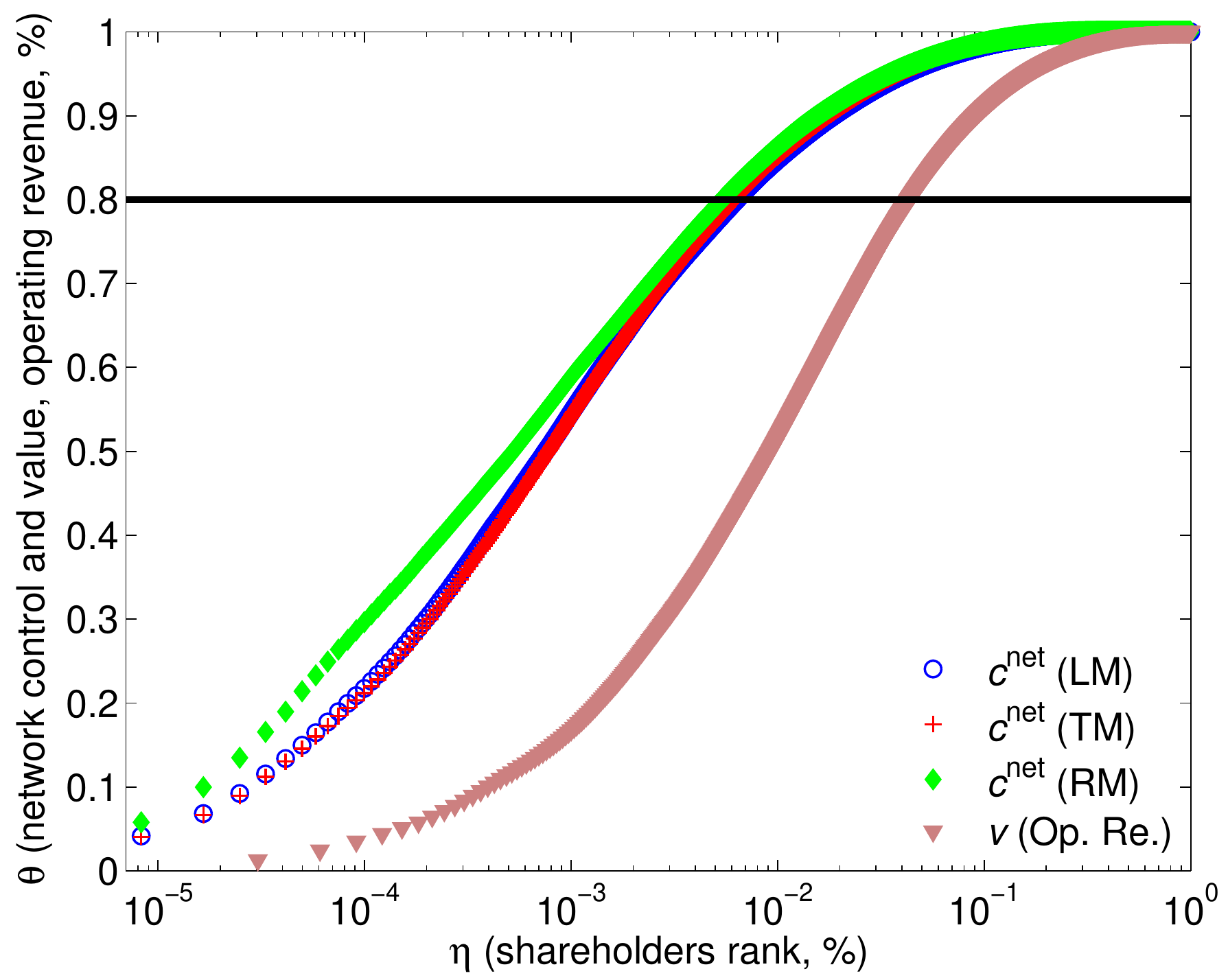}
  \caption{{\bf Concentration of network control and operating
      revenue.} Economic actors (TNCs and shareholders) are sorted by
    descending importance, as given by $c^{\textrm {net}}$. A data
    point located at ($\eta,\theta$) corresponds to a fraction $\eta$
    of top economic actors cumulatively holding the fraction $\theta$
    of network control, value or operating revenue.  The different
    curves refer to network control computed with three models (LM,
    TM, RM), see SI Appendix, Sec. 3.1, and operating revenue. The
    horizontal line denotes a value of $\theta$ equal to $80\%$. The
    level of concentration is determined by the $\eta$ value of the
    intersection between each curve and the horizontal line. The scale
    is semi-log.}
\label{fig:concentration}
\end{center}
\end{figure}

It should be noticed that, although scholars have long measured the
concentration of wealth and income
\cite{atkinson2000handbook}, there is no prior quantitative estimation
for control. Constructing a Lorenz-like curve (Fig.\
\ref{fig:concentration}) allows one to identify the fraction
$\eta^\ast$ of top holders holding cumulatively $80\%$ of the total
network control. Thus, the smaller this fraction, the higher the
concentration. In principle, one could expect inequality of control to
be comparable to inequality of income across households and firms,
since shares of most corporations are publicly accessible in stock
markets. In contrast, we find that only $737$ top holders accumulate
$80\%$ of the control over the value of all TNCs (see also the list of
the top $50$ holders in Tbl. S1 of SI Appendix, Sec. 8.3). The
corresponding level of concentration is $\eta_1^\ast=0.61\%$, to be
compared with $\eta^\ast_2=4.35\%$ for operating revenue. Other
sensible comparisons include: income distribution in developed
countries with $\eta^\ast_3
\sim 5\%-10\%$ \cite{atkinson2000handbook} and corporate revenue in
Fortune1000 ($\eta_4^\ast \sim 30 \%$ in 2009). This means that
network control is much more unequally distributed than wealth. In
particular, the top ranked actors hold a control ten times bigger than
what could be expected based on their wealth. The results are robust
with respect to the models used to estimate control, see Fig.\
\ref{fig:concentration} and Tbls.\ S2, S3.

\section*{Discussion}

The fact that control is highly concentrated in the hands of few top
holders does not determine if and how they are interconnected. It is
only by combining topology with control ranking that we obtain a full
characterization of the structure of control. A first question we are
now able to answer is where the top actors are located in the bow-tie.
As the reader may by now suspect, powerful actors tend to belong to
the core. In fact, the location of a TNC in the network does matter.
For instance, a randomly chosen TNC in the core has about $50\%$
chance of also being among the top holders, compared to, e.g., $6\%$
for the in-section (Tbl. S4). A second question concerns what share of
total control each component of the bow-tie holds. We find that,
despite its small size, the core holds collectively a large fraction
of the total network control. In detail, nearly $4/10$ of the control
over the economic value of TNCs in the world is held, via a
complicated web of ownership relations, by a group of $147$ TNCs in
the core, which has almost full control over itself. The top holders
within the core can thus be thought of as an economic ``super-entity''
in the global network of corporations. A relevant additional fact at
this point is that $3/4$ of the core are financial intermediaries.
Fig. \ref{fig:top} D shows a small subset of well-known financial
players and their links, providing an idea of the level of
entanglement of the entire core.

This remarkable finding raises at least two questions that are
fundamental to the understanding of the functioning of the global
economy. Firstly, what are the implication for global financial
stability? It is known that financial institutions establish financial
contracts, such as lending or credit derivatives, with several other
institutions. This allows them to diversify risk, but, at the same
time, it also exposes them to contagion
\cite{allen2000financial}. Unfortunately, information on these
contracts is usually not disclosed due to strategic reasons. However,
in various countries, the existence of such financial ties is
correlated with the existence of ownership relations
\cite{santos2006american}. Thus, in the hypothesis that the structure
of the ownership network is a good proxy for that of the financial
network, this implies that the global financial network is also very
intricate. Recent works have shown that when a financial network is
very densely connected it is prone to systemic risk
\cite{battiston.ea09,stiglitz2010risk}. Indeed, while in good times
the network is seemingly robust, in bad times firms go into distress
simultaneously. This \textit{knife-edge} property
\cite{alessandri2009,may2008ecology} was witnessed during the recent
financial turmoil.

Secondly, what are the implications for market competition? Since many
TNCs in the core have overlapping domains of activity, the fact that
they are connected by ownership relations could facilitate the
formation of blocs, which would hamper market competition
\cite{gilo.ea06}. Remarkably, the existence of such a core in the
global market was never documented before and thus, so far, no
scientific study demonstrates or excludes that this international
``super-entity'' has ever acted as a bloc. However, some examples
suggest that this is not an unlikely scenario. For instance, previous
studies have shown how even small cross-shareholding structures, at a
national level, can affect market competition in sectors such as
airline, automobile and steel, as well as the financial one
\cite{gilo.ea06,obrien.ea99}. At the same time, antitrust institutions
around the world (e.g., the UK Office of Fair Trade) closely monitor
complex ownership structures within their national borders. The fact
that international data sets as well as methods to handle large
networks became available only very recently, may explain how this
finding could go unnoticed for so long.

Two issues are worth being addressed here. One may question the idea
of putting together data of ownership across countries with diverse
legal settings. However, previous empirical work shows that of all
possible determinants affecting ownership relations in different
countries (e.g., tax rules, level of corruption, institutional
settings, etc.), only the level of investor protection is
statistically relevant \cite{laporta.ea99}. In any case, it is
remarkable that our results on concentration are robust with respect
to three very different models used to infer control from ownership. The
second issue concerns the control that financial institutions
effectively exert. According to some theoretical arguments, in
general, financial institutions do not invest in equity shares in
order to exert control. However, there is also empirical evidence of
the opposite \cite[SI Appendix, Sec.  8.1]{santos2006american}. Our
results show that, globally, top holders are at least in the position
to exert considerable control, either formally (e.g., voting in
shareholder and board meetings) or via informal negotiations.

Beyond the relevance of these results for economics and policy making,
our methodology can be applied to identify key nodes in any real-world
network in which a scalar quantity (e.g., resources or energy) flows
along directed weighted links. From an empirical point of view, a
bow-tie structure with a very small and influential core is a new
observation in the study of complex networks. We conjecture that it
may be present in other types of networks where ``rich-get-richer''
mechanisms are at work (although a degree preferential-attachment
\cite{Barabasi1999} alone does not produce a bow-tie). However, the
fact that the core is so densely connected could be seen as a
generalization of the ``rich-club phenomenon'' with control in the
role of degree \cite[SI Appendix, Sec.
8.2]{colizza.ea06,fagiolo.ea09}. These related open issues could be
possibly understood by introducing control in a ``fitness model''
\cite{garlaschelli2007self} of network evolution.

\section*{Acknowledgments}
We acknowledge F. Schweitzer and C. Tessone for valuable feedback,
D. Garcia for generating the 3D figures, and the program Cuttlefish
used for networks layout.
\\
Authors acknowledge the financial support from: the ETH Competence
Center ``Coping with Crises in Complex Socio-Economic Systems'' (CCSS)
through ETH Research Grant CH1-01-08-2; the European Commission FP7
FET Open Project ``FOC'' No. 255987.

\clearpage

\renewcommand{\thefigure}{S\arabic{figure}}
\renewcommand{\thetable}{S\arabic{table}}
 
\setcounter{table}{0}
\setcounter{figure}{0}

\makeatletter
\renewcommand{\@biblabel}[1]{\quad#1.}
\makeatother

\title{Supporting Information: \\
The Network of Global Corporate Control}
\date{}
\author{Stefania Vitali, James B. Glattfelder and Stefano Battiston\\
Chair of Systems Design, ETH  Zurich, Kreuzplatz 5, 8032 Zurich, Switzerland}

\maketitle
\tableofcontents

\clearpage

\section{Acronyms and Abbreviations}
\label{sec:acronyms-abbr}

The list of acronyms and abbreviations used in the main text and this Supporting Online Material:\\ \\
BFS: breadth-first search (search algorithm)\\
CC: (weakly) connected component\\
FS: financial sector\\
IN: in-section of a bow-tie\\
LCC: largest CC\\
LM: linear model (for estimating control from ownership; see also RM and TM)\\
NACE: (industry standard classification system )\\
OCC: other connected components (everything outside the LCC)\\
OECD: Organization for Economic Co-operation and Development\\
OR: operating revenue\\
OUT: out-section of a bow-tie\\
PC: participated company\\
RM: relative model (for estimating control from ownership; see also LM and TM)\\
SCC: strongly connected component (in the main text, this is synonymous with the core of the bow-tie in the LCC)\\
SH: shareholder (economic actors holding shares in TNCs)\\
TCH: top control-holder (list of TNCs and SHs that together hold 80\% of the network control)\\
TM: threshold model (for estimating control from ownership; see also LM and RM)\\
TNC: transnational corporation (OECD definition)\\
T\&T: tubes and tendrils (sections in a bow-tie that either connect IN
and OUT, are outgoing from IN, or ingoing to OUT, respectively)

\clearpage

\section{Data and TNC Network Detection}
\label{sec:data-tnc-network}

The Orbis 2007 marketing database\footnote{URL:
  \texttt{http://www.bvdep.com/en/ORBIS}. \label{foot}} comprises about 37 million
economic actors, both physical persons and firms located in 194
countries, and roughly 13 million directed and weighted ownership
links (equity relations). Among many others, information on the
industrial classification, geographical position and operating revenue
of the actors are provided. This data set is intended to track control
relationships rather than patrimonial relationships. Whenever
available, the percentage of ownership refers to shares associated
with voting rights.

The definition of TNCs given by the OECD\cite{OECD00} states that they
\begin{quote}
  \textit{[...] comprise companies and other entities established in more
    than one country and so linked that they may coordinate their
    operations in various ways, while one or more of these entities may
    be able to exercise a {\bf \it significant influence} over the
    activities of others, their degree of autonomy within the enterprise
    may vary widely from one multinational enterprise to
    another. Ownership may be private, state or mixed.}
\end{quote}
Accordingly, we select those companies which hold at least 10\% of
shares in companies located in more than one country. However, many
subsidiaries of large TNCs fulfill themselves this definition of TNCs
(e.g. The Coca-Cola Company owns Coca-Cola Hellenic Bottling Company
which in turn owns Coca-Cola Beverages Austria). Since for each
multinational group we are interested in retaining only one
representative, we exclude from the selection the companies for which
the so-called ultimate owner (i.e., the owner with the highest share
at each degree of ownership upstream of a
company\textsuperscript{\ref{foot}}) is quoted in a the stock market.
In substitution, we add the quoted ultimate owner to the list (if not
already included). In the example above, this procedure identifies
only the Coca-Cola Company as a TNC. Overall we obtain a list of 43060
TNCs located in 116 different countries, with 5675 TNCs quoted in
stock markets.

\begin{figure} 
  \vspace{-1em} 
 \begin{minipage}{0.9\textwidth}
\flushleft
    \textsf{\textbf{A}}\\
  \vspace{-6em} 
\centering 
\includegraphics[scale=0.425]{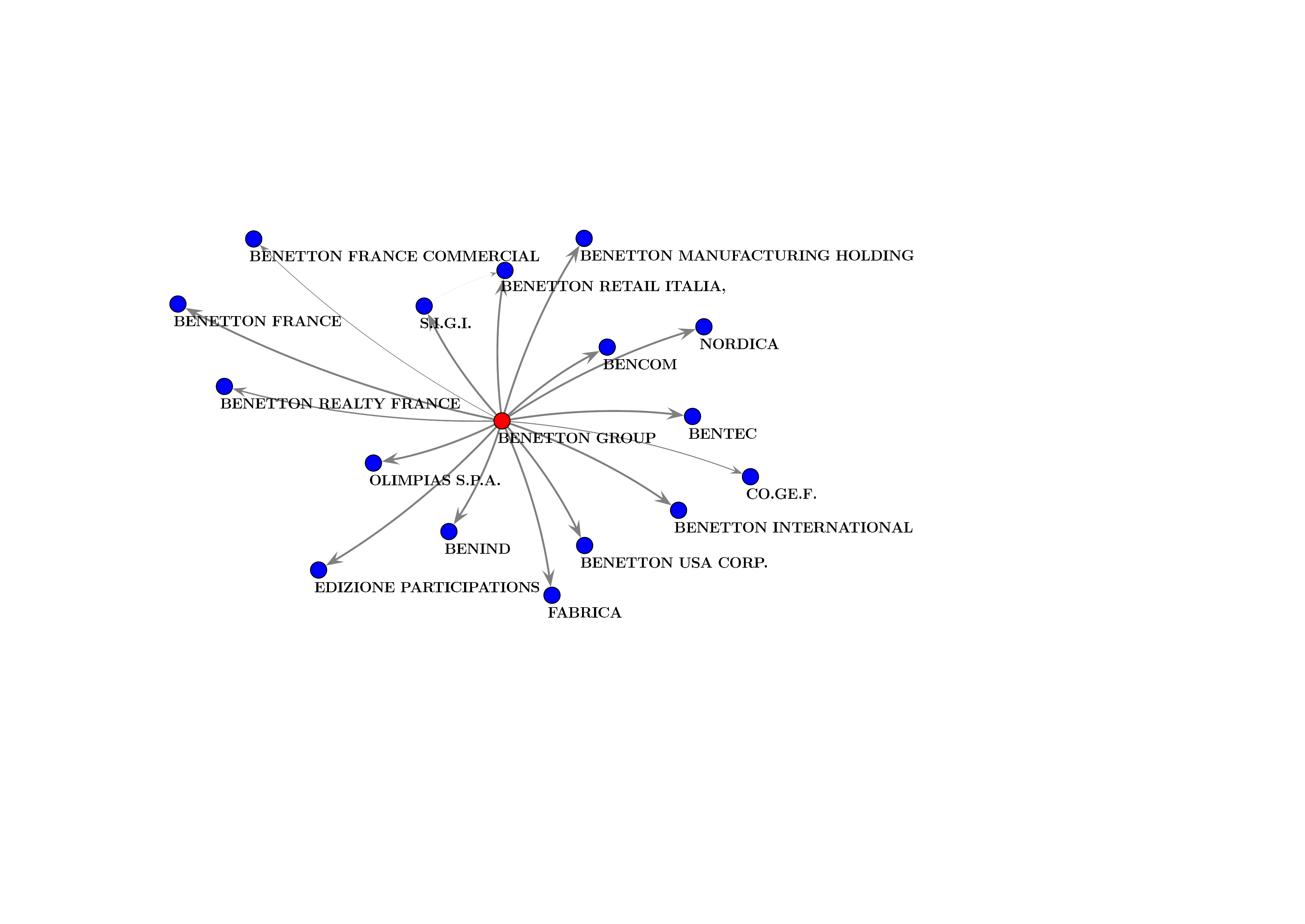}
\hspace{-30em}
  \vspace{-7em} 
  \end{minipage} \\
  \vspace{-6em} 
 \begin{minipage}{0.9\textwidth}
\flushleft
    \textsf{\textbf{B}}\\
  \vspace{-6em} 
\centering
\includegraphics[scale=0.425]{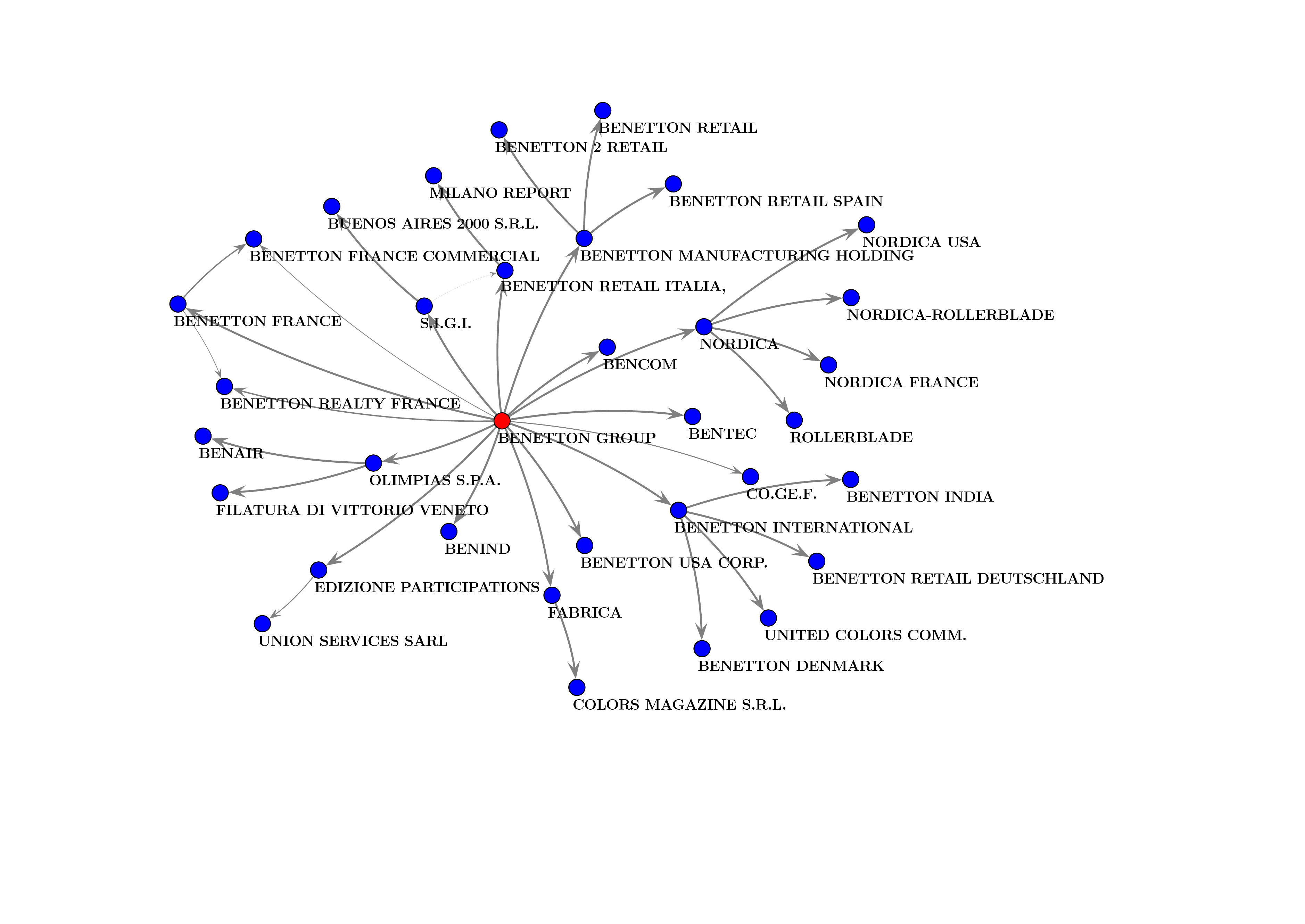}
  \end{minipage}
\caption{Illustration of the first two steps in
  the recursive exploration downstream of a TNC. Starting from ``Benetton
  Group'' the BFS explores all the direct neighbors ({\bf{\textsf A}}),
  and then the neighbors' neighbors ({\bf{\textsf B}}).} \label{sfig:bfs}
\end{figure}

\begin{figure}[ht]
  \begin{center}
  \includegraphics[width=0.7\textwidth]{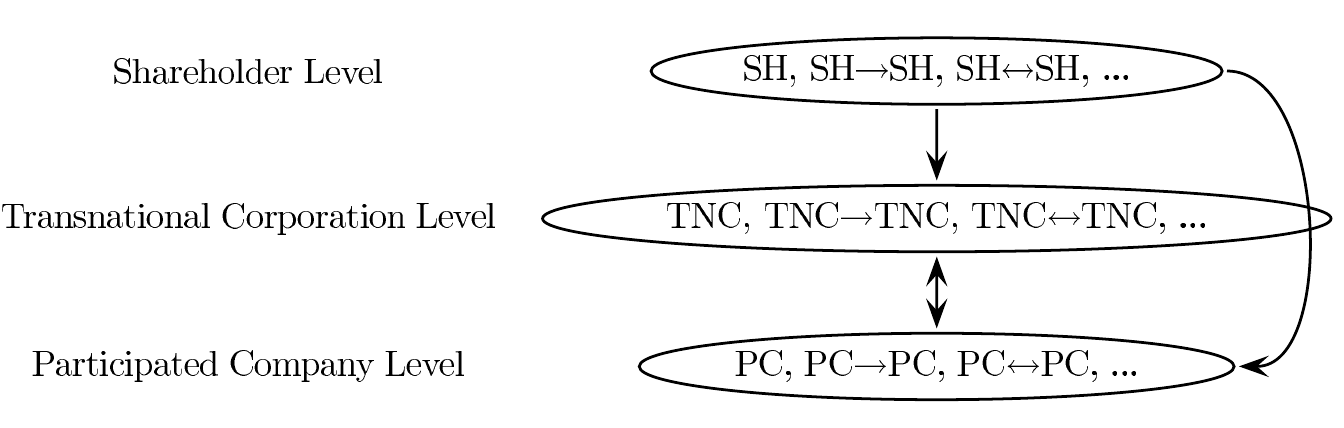}\\
\end{center}
\caption{ General structure of the TNC network.
  Three types of economic actors appear: 77456 SHs, 43060
  TNCs and 479992
  PCs. The network contains in total 600508 nodes, and 1006987
  links. Links are mainly from the TNCs to the PCs and amongst the PCs
  themselves.} \label{sfig:levels} 
\end{figure}

Starting from the list of TNCs, we explore recursively the neighborhood
of companies in the whole database. First, we proceed downstream of the
TNCs (see Fig. \ref{sfig:bfs}) with a breadth-first search (BFS) and we
identify all companies participated directly and indirectly by the
TNCs. We then proceed in a similar way upstream identifying all direct
and indirect shareholders of the TNCs. The resulting network can be
divided into three classes of nodes, TNC, SH and PC, as shown in
Fig. \ref{sfig:levels}. The TNC network constructed in this way consists
of 600508 economic entities and 1006987 corporate relations. Notice that
it may be possible to reach a PC from several TNCs, or to reach a TNC
from several SHs. In other words, paths proceeding downstream or upstream
of the TNCs may overlap, giving rise to CCs of various sizes.

It is worthwhile to distinguish the data set constructed here from the
one analysed in 
\cite{GlattfelderBattiston2009BackboneComplexNetworkssom}, 
which was not obtained using a recursive search, but with the simple
method of collecting only listed companies and their direct
shareholders. This method neglects all indirect paths involving
non-listed companies, so that the true ownership network was only
approximated. Moreover, 48 countries were analysed separately,
ignoring all cross-country links, an approach which inevitably leaves
out entirely the global structure of ownership. The aim there was to
construct disjoint national stock market networks, from which the
backbones were extracted and analyzed. Here, however, we focus on the
entire global topology.

\clearpage

\clearpage

\section{Network Control}

\label{sec:network-control}
In this section, we first recapitulate in detail the existing method for
computing the value or control in a network. In a second step, we
highlight two problems that plague this approach, especially in networks
with bow-tie topology (see main text, Sec. Network Topology).  The first
is that the control assigned to firms that are part of cross-shareholding
structures is overestimated. The second is a similar overestimation of
the control of the shareholders who are themselves not owned by
others. These two problems require independent solutions. In particular,
the second problem was never raised before in the literature. We provide
a novel algorithm that, for the first time, solves both problems and
allows the computation of control also for large networks. This method
represents a fundamental improvement to previous works, including our own
one \cite{GlattfelderBattiston2009BackboneComplexNetworkssom}, as
explained below in details. Finally, we illustrate the problem and the
corrections introduced by the algorithm using a representative example of
a small bow-tie network.

\subsection{The Existing Methodology}
\label{sec:existing-methodology}

While ownership is an objective quantity given by the percentage of
shares owned in a company, control, reflected in voting rights, can only
be estimated using a model. There are two steps involved in the
derivation of the notion of control we use in this work. Firstly, direct
control is estimated from the direct ownership relations. Network control
is then computed on the basis of direct control considering all paths in
the network.

For the computation of the direct control, we use three models: the
linear model, applying the one-share-one-vote rule \cite{goergen.ea05som,
  deminor_rating05}, the threshold model \cite{laporta.ea99som} and the
relative control model
\cite{GlattfelderBattiston2009BackboneComplexNetworkssom}. In the main
part of the text, we denote these three models as LM, TM and RM,
respectively. According to the LM, there is no deviation between
ownership and control, thus the direct control matrix coincides with the
ownership matrix, $L_{ij} = W_{ij}$. In the TM, full control over a
company is assigned to the actor holding a number of shares higher than a
predefined threshold (50\% in our case), while the other holders are
assigned zero control. The control matrix for the threshold model is
denoted as $T_{ij}$. Finally, the RM assigns control based on the
relative fraction of ownership shares that each shareholder has (using a
Herfindhal-like concentration index). The control matrix is defined as
$R_{ij} := W_{ij}^{2} / (\sum_{l=1}^{k^{in}_j} W_{lj}^{2})$. In
particular, the RM assigns high control to a shareholder with a small
share in absolute terms, if this share is significantly bigger than the
shares of all the other shareholders. For each of these three control
matrices, network control is computed with the same procedure. In the
main text we use the TM as our main measure, and compare all the results
with the LM and the RM. It should be stressed that the global findings
are insensitive to the chosen model of direct control.

As explained in the main text, the value of the portfolio of firms owned
directly by $i$ should be computed taking into account the value of the
firms owned by the firms in the portfolio and so on. Thus, the network
portfolio value $p^{\textrm{net}}_i$ consists of the value gained
indirectly plus the value of the direct portfolio: $p^{\textrm{net}}_i =
\sum_jW_{ij}v_j + \sum_jW_{ij}p^{\textrm{net}}_j$. The vector $v$
represents the intrinsic value of the firms (e.g., operating revenue,
total assets or market capitalization). Here we use operating revenue,
because it is readily available for the economic actors under
investigation and it is comparable across sectors (this is not true for
total assets). In analogy to the definition above, we introduce the
\textit{network control (value)} 
\cite{GlattfelderBattiston2009BackboneComplexNetworkssom}.
This quantity measures the value controlled by a shareholder taking
into account the network of firms in which it has direct or indirect
shares. In matrix notation,
\begin{equation}
\label{eq:controlcentrality}
c^{\textrm{net}}  = \mathcal{C}c^{\textrm{net}}+\mathcal{C}v, 
\end{equation}
where $\mathcal{C} \in \{L, T, R\}$ is one of the three direct control
matrices. The solution to Eq. (\ref{eq:controlcentrality}) is given by
\begin{equation}
\label{eq:netwcontrol}
c^{\textrm{net}} =  (I-\mathcal{C} )^{-1}  \mathcal{C} v  =: \tilde{\mathcal{C}} v.
\end{equation}
For the matrix $(I-\mathcal{C})$ to be non-negative and non-singular, a
sufficient condition is that the Frobenius root of $\mathcal{C}$ is
smaller than one, \mbox{$\lambda(\mathcal{C})<1$}. This is ensured by the
following requirement: in each strongly connected component $\mathcal{S}$
there exists at least one node $j$ such that $\sum_{i\in \mathcal{S}}
\mathcal{C}_{ij}<1$. This means that there exists no subset of $k$ firms $(k =
1,\dots , n)$ that are entirely controlled by the $k$ firms themselves, a
condition which is always fulfilled.

By taking the series expansion of $(I-\mathcal{C})^{-1}$, it can be
proven that: $ \mathcal{C} (I-\mathcal{C})^{-1} = (I-\mathcal{C})^{-1}
\mathcal{C}$. As a consequence, $\tilde{\mathcal{C}}$ in
Eq. (\ref{eq:netwcontrol}) coincides with the solution to the equation
\begin{equation}
\label{eq:Ctildeeq}
\tilde{\mathcal{C}}_{ij} = \mathcal{C}_{ij} + \sum_{k} \tilde{\mathcal{C}}_{ik} \mathcal{C}_{kj}.
\end{equation}
This is corresponds to the definition of integrated ownership given in
\cite{brioschi.ea89som}.  Hence, as in
\cite{GlattfelderBattiston2009BackboneComplexNetworkssom}, we can
interpret $c^{\textrm{net}}$ as the value of control an economic actor
gains from all its direct and indirect paths in the network.

Notice that Eq. (\ref{eq:controlcentrality}) is related to the
notion of eigenvector centrality used to investigate power and influence
both in social and economic networks
\cite{bonacich1987powersom,ballester.ea.2006whosom}. There is also 
an additional interpretation of network control in terms a physical
system in which a quantity is flowing along the links of the network
\cite{GlattfelderBattiston2009BackboneComplexNetworkssom}. In this
picture, nodes associated with a value $v_j$ produce $v_j$ units of the
quantity at time $t=1$. The weight of a link $ij$, given by the adjacency
matrix entry $A_{ij}$, determines the fraction of $v_j$ that flows
through it. Then the \textit{inflow}, i.e. the flow $\phi_i$ entering
the node $i$ from each node $j$ at time $t$ is the fraction $A_{ij}$ of
the quantity produced by $j$ plus the same fraction of the
inflow of $j$:
\begin{equation}
  \label{eq:mass-inflow}
  \phi_i(t+1) =  \sum_j A_{ij}  \phi_i(t) + \sum_j A_{ij} v_j ,
\end{equation}
 In matrix notation, at the steady state, this yields
\begin{equation}
  \label{eq:mass-inflow-matrix}
  \phi= A \phi + A v,
\end{equation}
which is formally identical to Eq.  (\ref{eq:controlcentrality}). Thus
if $v$ corresponds to an intrinsic economic value of the nodes, then
the network control corresponds to the inflow of control over this
value. The network portfolio value of a node is determined by the
total inflow of value entering the node.

Next to network control, a related quantity is the so-called
{\it network value}
\begin{equation}
\label{eq:netwvalue1}
v^{\textrm{net}} =  \mathcal{C} v^{\textrm{net}} +v, 
\end{equation}
which is akin to a Hubbell index centrality measure
\cite{hubbell65}. This measure is well-established in
the literature \cite{brioschi.ea89som}. The solution is
$v^{\textrm{net}} = (I-\mathcal{C})^{-1}v$.  By noting that
\begin{equation}
\label{eq:corresp}
\mathcal{C} v^{\textrm{net}}= \mathcal{C}(I-\mathcal{C})^{-1}v = \tilde{\mathcal{C}} v,
\end{equation}
we find
\begin{equation}
\label{eq:netwvalcontrl}
v^{\textrm{net}}  = \tilde{\mathcal{C}} v + v =  c^{\textrm{net}}  + v.
\end{equation}
In other words, the network value of an economic actor is given by its
intrinsic value plus the value gained from network control. It is an
estimate of the overall value a corporation has in an ownership
network. Notice that network value and network control of a company
can differ considerably. As an example, Wall Mart is in top rank by
operating revenue but it has no equity shares in other TNCs and thus
its network control is zero. In contrast, a small firm can acquire
enormous network control via shares in corporations with large
operating revenue.

From Eq. (\ref{eq:corresp}), where $c^{\textrm{net}} =
\tilde{\mathcal{C}} v = \mathcal{C} v^{\textrm{net}}$, network control
can either be understood as the value of control gained from the
intrinsic value reachable by all direct and indirect paths or the value
of control given by the network value of directly controlled firms.

\subsection{The Algorithm: Computing Control While Remedying the Problems}

Unfortunately, the equations defining network control and network
value suffer from three drawbacks. Firstly, the computation
overestimates control when there are cycles in the network (for
example in an SCC\footnote{For more information see SI
Sec. \ref{sec:scc}.}), i.e., when the number of inter-firm
cross-shareholdings grows
\cite{baldone.ea98som}. Secondly, as we have discovered, it also leads to paradoxical
situations. Consider for instance an SCC that is reachable from a
single root-node $r$ that owns an arbitrarily small share in one of
the firms in the SCC. The above definition assigns to such a node the
sum of the intrinsic value of all the nodes in the SCC. This is
obviously not a correct estimate of the control of the node $r$. These
two issues are best understood in the flow analogy. Indeed, in a dense
SCC control flows through the nodes many times.  The smaller the
incoming links from the IN are the longer it takes until the flow
stops, as, in the steady state, everything ultimately flows to and
accumulates in the root-nodes.  However, since control corresponds to
the total inflow over an infinite time this exaggerates the control of
the nodes in the SCC and all the control ultimately flows to the
root-nodes. Thirdly, for large networks, the computation of the
inverse matrix can be intractable.  Here, for the first time, we
overcome the aforementioned problems and propose a new methodology
that consists of applying an algorithm to compute network control by
treating different components of the network separately.

We first illustrate the algorithm for the computation of
$v^{\textrm{net}}$. Then $c^{\textrm{net}} = v^{\textrm{net}}-v$. In
order to calculate the network value for any specific node $i$, we
extract the whole subnetwork that is downstream of a node $i$,
including $i$. For this purpose, a breadth-first-search (BFS) returns
the set of all nodes reachable from $i$, going in the direction of the
links. Then, all the links among these nodes are obtained from the
control matrix of the entire network, except for the links pointing to
$i$ which are removed. This ensures that there are no cycles involving
$i$ present in the subnetwork. Let $B(i)$ denote the adjacency matrix
of such a subnetwork, including $i$, extracted from the control matrix
$\mathcal{C}= (L, T, R)$. Without loss of generality, we can relabel
the nodes so that $i=1$. Since node $1$ has now no incoming links, we
can decompose $B=B(1)$ as follows:
\begin{equation}
\label{eq:adjBFS}
B = 
\left(\begin{array}{c|c}
0 &d\\\hline
\vec 0 & B^{\textrm{sub}}
\end{array}\right),
\end{equation}
where $d$ is the row-vector of all links originating from node 1, and
$B^{\textrm{sub}}$ is associated with the subgraph of the nodes
downstream of $i$. The value of these nodes is given by the column-vector
$v^{\textrm{sub}}$.  By replacing the the matrix $B$ in the expression
$v^{\textrm{net}} = \tilde{\mathcal{C}} v +v =
\mathcal{C}(I-\mathcal{C})^{-1} v +v$ and taking the first component we
obtain:
\begin{eqnarray}
\label{eq:netwvalueBFS}
v^{\textrm{net}} (1) &=& \left[ B (I-B)^{-1} v \right]_1 + v_1  \notag \\
&=& d (I^{\textrm{sub}}-B^{\textrm{sub}})^{-1} v^{\textrm{sub}}  + v_1 =: 
\tilde d \cdot v^{\textrm{sub}}  + v_1, \label{eq:dtilde}
\end{eqnarray}
where now $c^{\textrm{net}} (1) := \tilde d \cdot v^{\textrm{sub}} = d (I^{\textrm{sub}}-B^{\textrm{sub}})^{-1} v^{\textrm{sub}}$.

Notice that if node $i$ has zero in-degree, this procedure yields the
same result as the previous formula: $\tilde{B}_{(i,*)} = (0, \tilde d )
= \tilde{\mathcal{C}}_{(i,*)}$. The notation $A_{(i,*)}$ for a matrix is
understood as taking its $i$-th row. In the next section it is shown that
our calculation is in fact equivalent to the correction proposed by
\cite{baldone.ea98som} to address the problems of the overestimation of
network value in the case of ownership due to the presence of cycles.

However, both methods still suffer from the problem of root nodes
accumulating all the control. This issue was previously overlooked
because the cases analysed did not have a bow-tie structure and because
the focus was not on the empirical analysis of control. To solve this
issue, we adjust our algorithm to pay special attention to the IN-nodes
of an SCC. We partition the bow-tie associated with this SCC into its
components: the IN (to which we also add the T\&T), the SCC itself, and
the OUT. Then, we proceed in multiple steps to compute the network value
for all parts in sequence. In this way, the control flows from the OUT,
via the SCC to the IN. Finally, the network control is computed from the
network value as $c^{\textrm{net}} = v^{\textrm{net}} - v$.  In detail,
our algorithm works as follows:
\begin{enumerate}
\item OUT: Compute the network value $v^{\textrm{net}}(i)$ for all the
  nodes in the OUT using Eq.  (\ref{eq:dtilde}).
\item OUT $\to$ SCC: Identify the subset $\mathcal{S}1$ of nodes in the SCC
  pointing to nodes in the OUT, the latter subset denoted as $\mathcal{O}$.
  To account for the control
  entering the SCC from the OUT, compute the network value of these
  selected nodes by applying $v^{\textrm{net}}(s) = \sum_o
  \mathcal{C}_{so} v^{\textrm{net}}(o)+ v_s$ to them.  This is an
  adaptation of Eq. \ref{eq:netwvalcontrl}, where $s$ and $o$ are labels
  of nodes in $\mathcal{S}1$ and $\mathcal{O}$, respectively.  Note that
  we only needed to consider the direct links for this.  This computation
  is also equivalent to applying Eq.  (\ref{eq:dtilde}), which considers
  the downstream subnetworks of $\mathcal{S}1$, i.e., the whole OUT.
\item SCC: Employ Eq. (\ref{eq:dtilde}) to the SCC-nodes restricting the
  BFS to retrieve only nodes in the SCC itself. Note that for those
  SCC-nodes that were already considered in step 2, their network
  value is now taken as the intrinsic value in the computation. This
  means one first needs to assign $v_i \mapsto v^{\textrm{net}} (i) +
  v_i$.
\item SCC $\to$ IN: In this step we solve the problem of the root-nodes
  acquiring an exaggerated fraction of the network value. For the subset of
  IN-nodes $\mathcal{I}$ directly connected to some SCC-nodes $\mathcal{S}2$,
  we again apply $v^{\textrm{net}}(i)=\sum_{s} \mathcal{C}_{is}
  v^{\textrm{net}}(s)+ v_i$, where $i$ and $s$ are labels of nodes in
  $\mathcal{I}$ and $\mathcal{S}2$, respectively. However, note that due
  to the cycles present in the SCC, this computation is not equivalent to
  Eq. (\ref{eq:dtilde}). In this way only the share of network value over
  the SCC which is not owned by other SCC-nodes is transferred to the
  IN-nodes.
\item IN: Finally, use Eq. (\ref{eq:dtilde}) for assigning the network
  value to the nodes in the IN-subnetwork. In this case the BFS should
  not consider the SCC-nodes since their value has been already
  transfer-ed to their first neighbors in the IN. However, it should
  retrieve the T\&T departing from the IN. Again, for the IN-nodes
  treated in step 4, first assign $v_i \mapsto v^{\textrm{net}} (i) +
  v_i$.
\end{enumerate}
Notice that if any part of the bow-tie structure contains additional
smaller SCCs, these should be treated first, by applying steps two to
four.

This dissection of the network into its bow-tie components also reduces
the computational problems. Although we perform a BFS for each node and
compute the inverse of the resulting adjacency matrix of the subnetwork
as seen in Eq. (\ref{eq:dtilde}), the smaller sizes of the subnetworks
allow faster computations.

\begin{figure}[ht]
  \begin{center}
  \includegraphics[width=0.45\textwidth,angle=0]{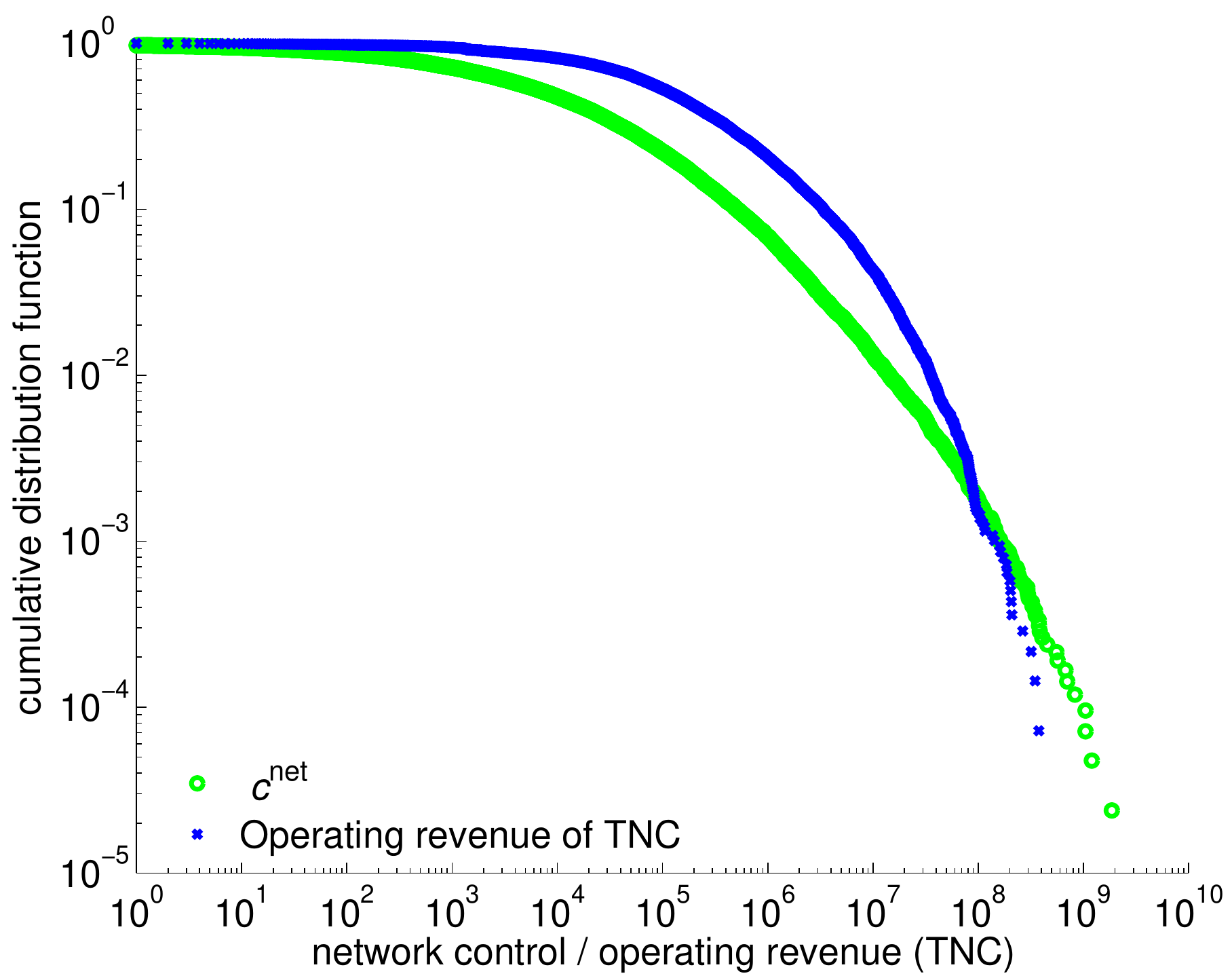}\\
\end{center}
\caption{Cumulative distribution function of network control and operating revenue.
The network control (TM) in the LCC and the operating revenue of the
TNCs in the LCC, from which it is computed, is shown.} \label{sfig:ornc}
\end{figure}

To summarize, using one of the three adjacency matrices estimating direct
control, $\mathcal{C} \in \{L, T, R\}$, we can compute the corresponding
network value for a corporation: $v^{\textrm{net}}_i$.  By deducting the
operating revenue, we retrieve the network control:
$c^{\textrm{net}}_i$. Operating revenue is taken for the value of the
TNCs ($v_i$). Fig. \ref{sfig:ornc} shows the distribution of the
operating revenue of the TNCs and the resulting network value.

\subsection{Proving the BFS Methodology Corrects for Cycles} 

Here we show that the BFS algorithm presented in the last section yields
an equivalent computation proposed in the literature to address the
problems of the presence of cycles leading to exaggerated network value.

In \cite{brioschi.ea89som} the notion of network value was introduced
based on ownership which corresponds, in the case of control, to
\begin{equation}
\label{eq:value_brioschi}
v^{\textrm{net}} = \tilde{\mathcal{C}} v + v,
\end{equation}
which in \cite{baldone.ea98som} was identified as being problematic. The
authors hence introduced a new model which overcomes this problem of
exaggerated indirect value in presence of cycles by introducing 
\begin{equation}
\label{eq:Ahatrec}
\hat {\mathcal{C}}_{ij} := {\mathcal{C}}_{ij} + \sum_{k \neq i} \hat {\mathcal{C}}_{ik}{\mathcal{C}}_{kj}.
\end{equation}
This means that the original matrix $\mathcal{C}$ defined in
Eq. (\ref{eq:Ctildeeq}) is corrected by removing all indirect
self-loops of any node $i$. If the network has no cycles, then Eqs. (\ref{eq:Ctildeeq})
and (\ref{eq:Ahatrec}) yield identical solutions.

We introduce here for the first time a correction operator, that
incorporates this modification and makes the associated computations
clearer
\begin{equation}
\label{eq:defd}
\mathcal{D} :=  \textrm{diag}((I- \mathcal{C})^{-1}) ^{-1} = I -  \textrm{diag}(\hat{\mathcal{C}}),
\end{equation}
where $\textrm{diag}(A)$ is the matrix of the diagonal of the matrix
$A$. It can be shown that
\begin{equation}
\label{eq:chatd}
\hat{\mathcal{C}} = \mathcal{D} \tilde{\mathcal{C}}.
\end{equation}
The associated corrected network value can be identified as 
\begin{equation}
\label{eq:value_baldone}
\hat v^{\textrm{net}} = \mathcal{D}  v^{\textrm{net}} = \hat{\mathcal{C}} v + \mathcal{D} v . 
\end{equation}

Our proposed methodology also corrects for cycles in an equivalent
way. This can be seen as follows. By applying the BFS algorithm to node
$i$, we extract the adjacency matrix $B(i)$ of the subnetwork of nodes
downstream. From Eq. (\ref{eq:Ahatrec}) it holds by construction that
\begin{equation}
\tilde B(i)_{ij} = \hat{\mathcal{C}}_{ij} - \hat{\mathcal{C}}_{ii}, 
\end{equation}
where $\tilde B(i)$ is defined equivalently to
Eq. (\ref{eq:netwcontrol}). In a more compact notation
\begin{equation}
\tilde B(i)_{i*} = \hat{\mathcal{C}}_{i*} - [\textrm{diag} ( \hat{\mathcal{C}})]_{i*}.
\end{equation}
Employing Eq.  (\ref{eq:defd}) we find that $\tilde B(i)_{i*} + I_{i*} =
\hat{\mathcal{C}}_{i*} + \mathcal{D}_{i*}$, or equivalently
\begin{eqnarray}
 &\hat{\mathcal{C}}_{i*} v + \mathcal{D}_{i*} v =  \mathcal{D}_{i*} (\tilde{\mathcal{C}}_{i*} v + v_i )
  = \mathcal{D}_{i*} v^{\textrm{net}}  =: \hat v^{\textrm{net}}_i \\
 &=  \tilde B_{i*}(i) v + v_i =  c^{\textrm{net}}(i) + v_i  =: v^{\textrm{net}} (i). \label{eq:our_nc}
\end{eqnarray}
This concludes that our BFS method and the results in
\cite{baldone.ea98som} are identical: $\hat v^{\textrm{net}}_i =  v^{\textrm{net}} (i)$.

\subsection{An Illustrated Example}

\begin{figure}[tH]
\centering
\vspace{-0.3cm}
\includegraphics[width=0.6\textwidth]{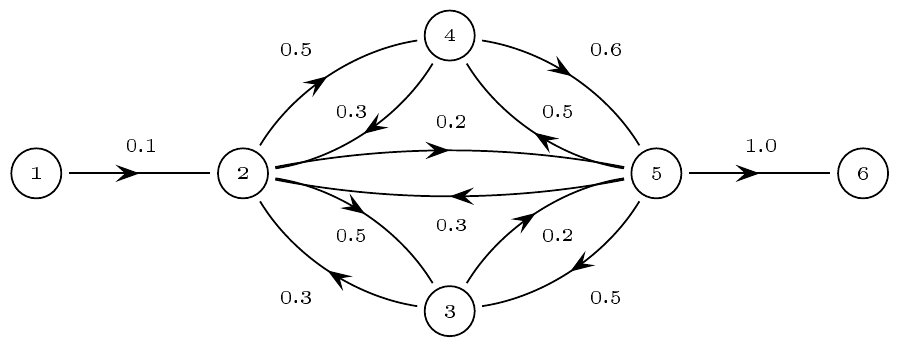}
\vspace{-0.0cm}
\caption{Simple bow-tie network topology.
Example with a high degree of
interconnectedness of the firms in the strongly connected
component (SCC).}\label{fig:netexb}
\end{figure}

Consider the network illustrated in Figure \ref{fig:netexb}. It is an
example of a simple bow-tie network topology. The SCC is constructed
in a way to highlight the problem of cross-shareholdings.  Hence there
are many cycles of indirect ownership originating and ending in each
firm in the core of the bow-tie.

We assume the underlying value of each firm to be one, i.e., $v=(1, 1,
1, 1, 1, 1)^t$, where $t$ denotes the transposition
operation. Moreover, we will employ the TM, hence
$\mathcal{C}_ij = W_{ij}$. This
results in the network value and the integrated value to be
\begin{equation}
v^{\textrm{net}} = \left( 
\begin{array}{r}
6 \\
50 \\
27  \\
49 \\
55 \\
1
\end{array} \right),
\end{equation}
using Eq. (\ref{eq:netwvalcontrl}).

So although the total value present in the network is $6 = \sum_i
v_i$, firm 5 has an disproportionately large network control of
$v^{\textrm{net}}_5 = 54$, highlighting the problem of overestimating
the control in the presence of cycles.

Employing the corrections proposed in \cite{baldone.ea98som}, i.e. by
computing the correction operator defined in Eq. (\ref{eq:defd}), one finds
\begin{equation}
\label{eq:exbd}
\mathcal{D} = \left(
\begin{matrix}
  1.000 & 0 & 0 & 0 & 0 & 0 \\
  0 & 0.100 & 0 & 0 & 0 & 0 \\
  0 & 0 & 0.162 & 0 & 0 & 0 \\
  0 & 0 & 0 & 0.095 & 0 & 0 \\
  0 & 0 & 0 & 0 & 0.086 & 0 \\
  0 & 0 & 0 & 0 & 0 & 1.000 \\
\end{matrix} \right) .
\end{equation}
From this, the corrected values can be computed from Eq. (\ref{eq:value_baldone})
\begin{equation}
\hat v^{\textrm{net}} = \left( 
\begin{array}{r}
6.000 \\
5.000 \\
4.378  \\
4.667 \\
4.714\\
1.000
\end{array} \right) .
\end{equation}
The correction reduces the values of the firms in the core of the
bow-tie by approximately one order of magnitude. This confirms that
$\hat v^{\textrm{net}}$ and $\hat c^{\textrm{net}}$ are indeed the
right measures to consider in the presence of SCCs in
the network.

Unfortunately, this example also highlights the second problem of the
methodology. It is clear, that root nodes accumulating all the
control. As mentioned, our proposed algorithm remedies this problem
while still correcting for the overestimation in cycles. One finds from
Eq. (\ref{eq:netwvalueBFS}) that
\begin{equation}
\left( 
\begin{array}{r}
v^{\textrm{net}} (1) \\
v^{\textrm{net}} (2) \\
v^{\textrm{net}} (3) \\
v^{\textrm{net}} (4) \\
v^{\textrm{net}} (5) \\
v^{\textrm{net}} (6) 
\end{array} \right)
= \left( 
\begin{array}{r}
  1.500\\
    5.000\\
    4.378\\
    4.667\\
    4.714\\
    1.000
\end{array} \right) ,
\end{equation}
illustrating the change from $v^{\textrm{net}}_1 = \hat
v^{\textrm{net}}_1 = 6 \ge v^{\textrm{net}} (1) = 1.5$. 

To summarize, employing $v^{\textrm{net}}$ for the computation of
control in networks with bow-tie topology overestimates the level of
control in the SCC by construction.  Using $\hat v^{\textrm{net}}$ on
the other hand always assigns the root nodes the highest control. Only
the measure $v^{\textrm{net}} (.)$ puts root and SCC-nodes on par with
each other and the leaf-nodes, allowing for the first time an accurate
analysis of the control of each node in the network.

\subsection{Relations To Previous Work}

To summarize, the relation the existing work is as follows. The notion of network
value\footnote{Although the authors only considered the case of ownership
  and not that of control, their methods are equivalent to the definition
  of control employing the LM.} was introduced in
\cite{brioschi.ea89som}, in addition to the integrated ownership
matrix. This matrix was later corrected in \cite{baldone.ea98som}. 

The
notion of network control was first defined in
\cite{GlattfelderBattiston2009BackboneComplexNetworkssom} without any of
the corrections described above. Because the networks analysed there
comprised only listed companies and their direct shareholders, it was
sufficient to apply the uncorrected methodology due to the absence of
long indirect paths, see SI Sec. \ref{sec:network-control}. In
contrast, in the present work, the full-fledged methodology with all
the corrections is required in order to consistently compute the flow
of control. This resulted in the introduction of the correction
operator and its application to the network value and network
control. As mentioned, this allowed us to identify a second problem
with the methodology. Subsequently, we have incorporated these
insights into an algorithm that is suitable for large networks,
correcting all potential problems with computing control. Finally, we
also uncover the relationship between network control and network
value.

\clearpage

\section{Degree and Strength Distribution Analysis}

\begin{figure}[b]
\begin{center}
  \begin{minipage}{0.45\textwidth}
    \begin{center}
      {\bf \textsf A} \includegraphics[scale=0.377,angle=0]{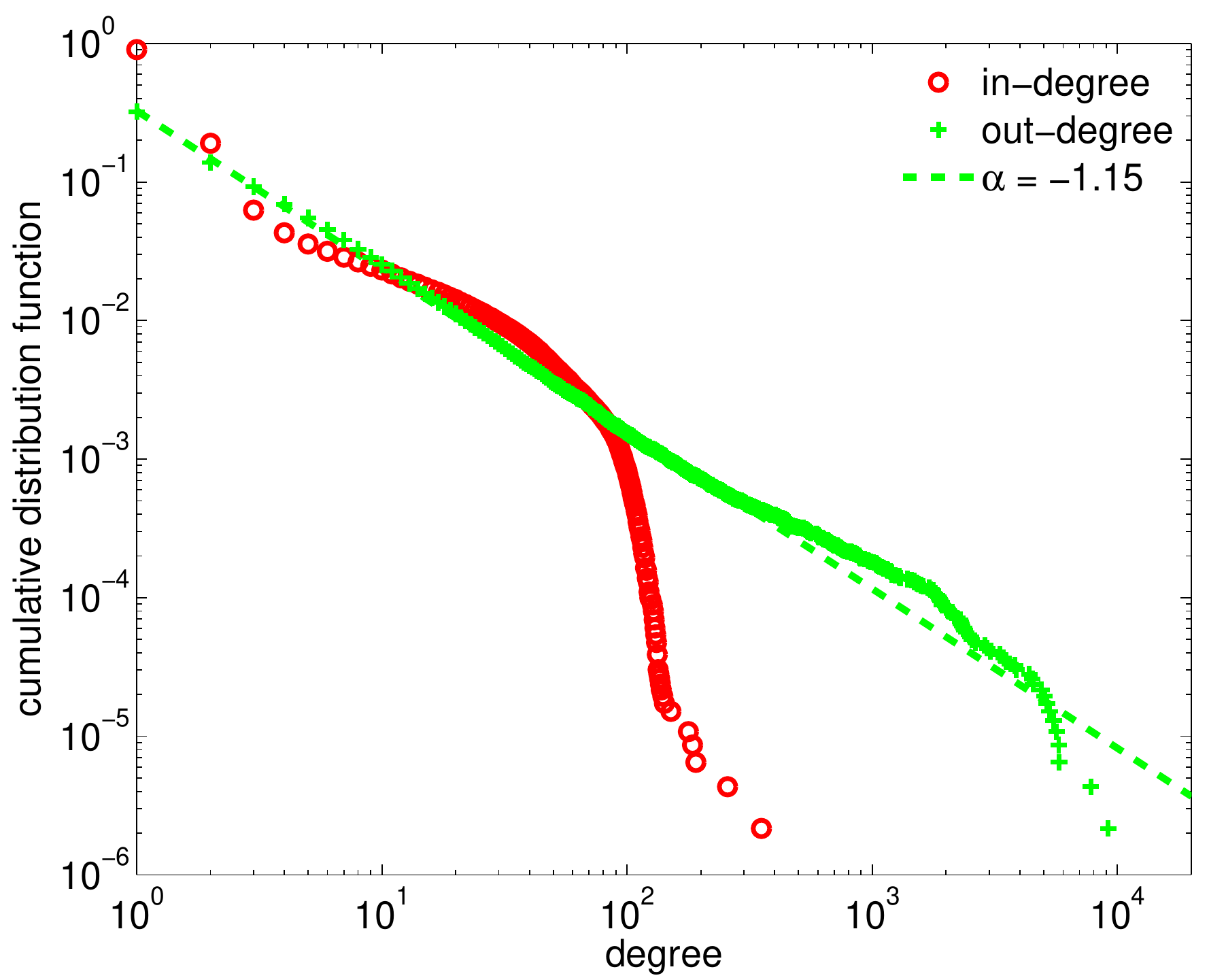}
    \end{center}
  \end{minipage}
  \begin{minipage}{0.45\textwidth}
    \begin{center}
      {\bf \textsf B} \includegraphics[scale=0.377,angle=0]{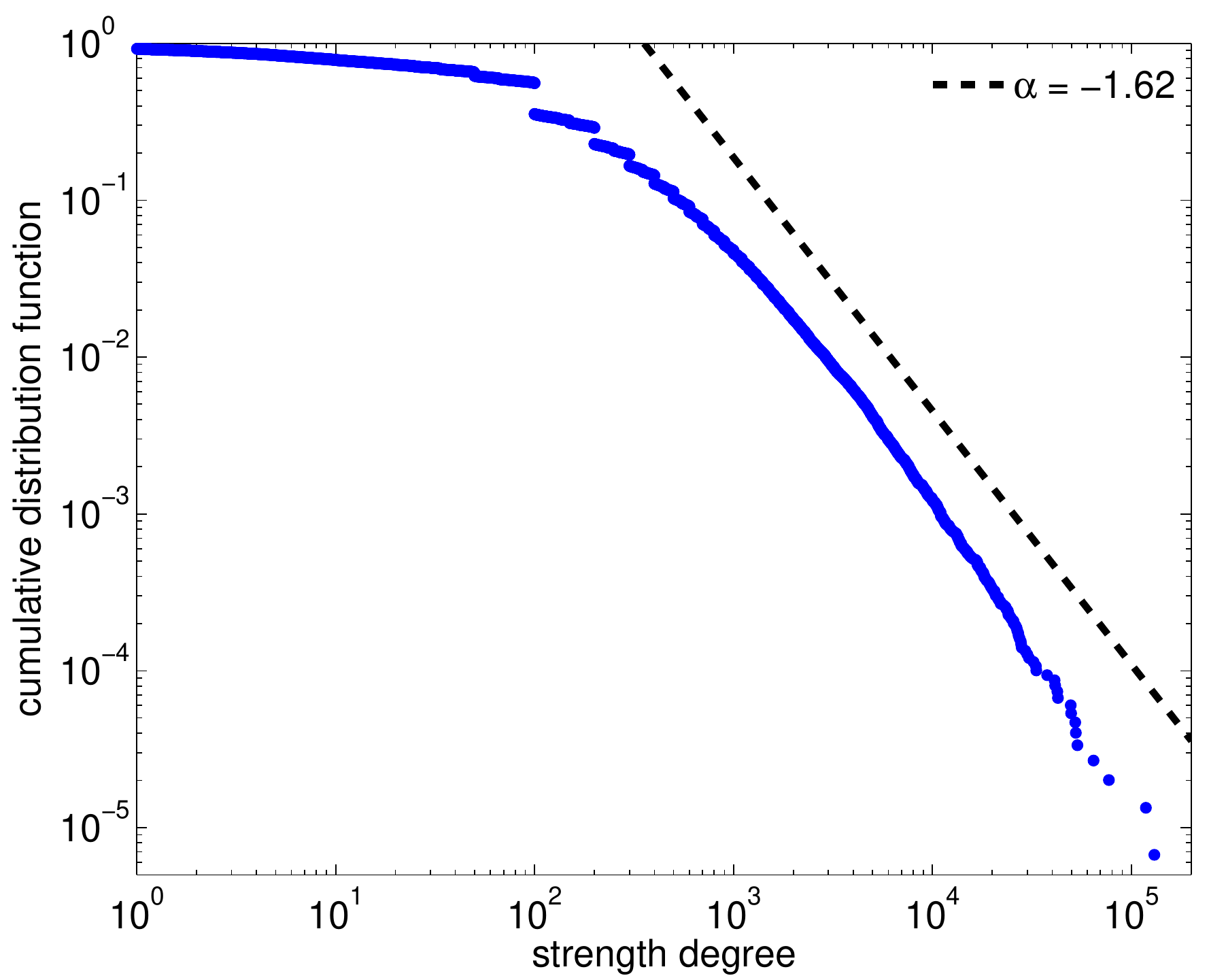}
        \end{center}
  \end{minipage}
\end{center}
\caption{Various distribution functions.
 ({\bf{\textsf A}}) Cumulative distribution function of the in- and
 out-degree of the nodes in the LCC (log-log scale). The power-law
 exponent for the corresponding probability density function of the
 out-degree is estimated to be -2.15. ({\bf{\textsf B}}) Cumulative
 distribution function of the node strength in the LCC (log-log
 scale). As a reference, a power-law with an exponent of $-1.62$ is
 displayed.} \label{sfig:distro}
\end{figure}

The study of the node degree refers to the distribution of the number of
in-going and out-going relations. The number of outgoing links of a node
corresponds to the number of firms in which a shareholder owns shares.
It is a rough measure of the portfolio diversification.  The in-degree
corresponds to the number of shareholders owning shares in a given
firm. It can be thought of as a proxy for control fragmentation. In the
TNC network, the out-degree can be approximated by a power law
distribution with the exponent -2.15 (see Fig.  \ref{sfig:distro}A). The
majority of the economic actors points to few others resulting in a low
out-degree. At the same time, there are a few nodes with a very high
out-degree (the maximum number of companies owned by a single economic
actor exceeds 5000 for some financial companies). On the other hand, the
in-degree distribution, i.e., the number of shareholders of a company,
behaves differently: the frequency of nodes with high in-degree decreases
very fast. This is due to the fact that the database cannot provide all
the shareholders of a company, especially those that hold only very small
shares.

Next to the study of the node degree, we also investigate the strength
which is defined as $\sum_j W_{ij}$, that is, the sum of all the weighed
participations a company $i$ has in other companies $j$ (see
Fig. \ref{sfig:distro}B). It is a measure of the weight connectivity and
gives information on how strong the ownership relationships of each node
are.

\clearpage

\section{Connected Component Analysis}

\begin{figure}[b]
   \begin{center}
   \includegraphics[scale=0.4,angle=0]{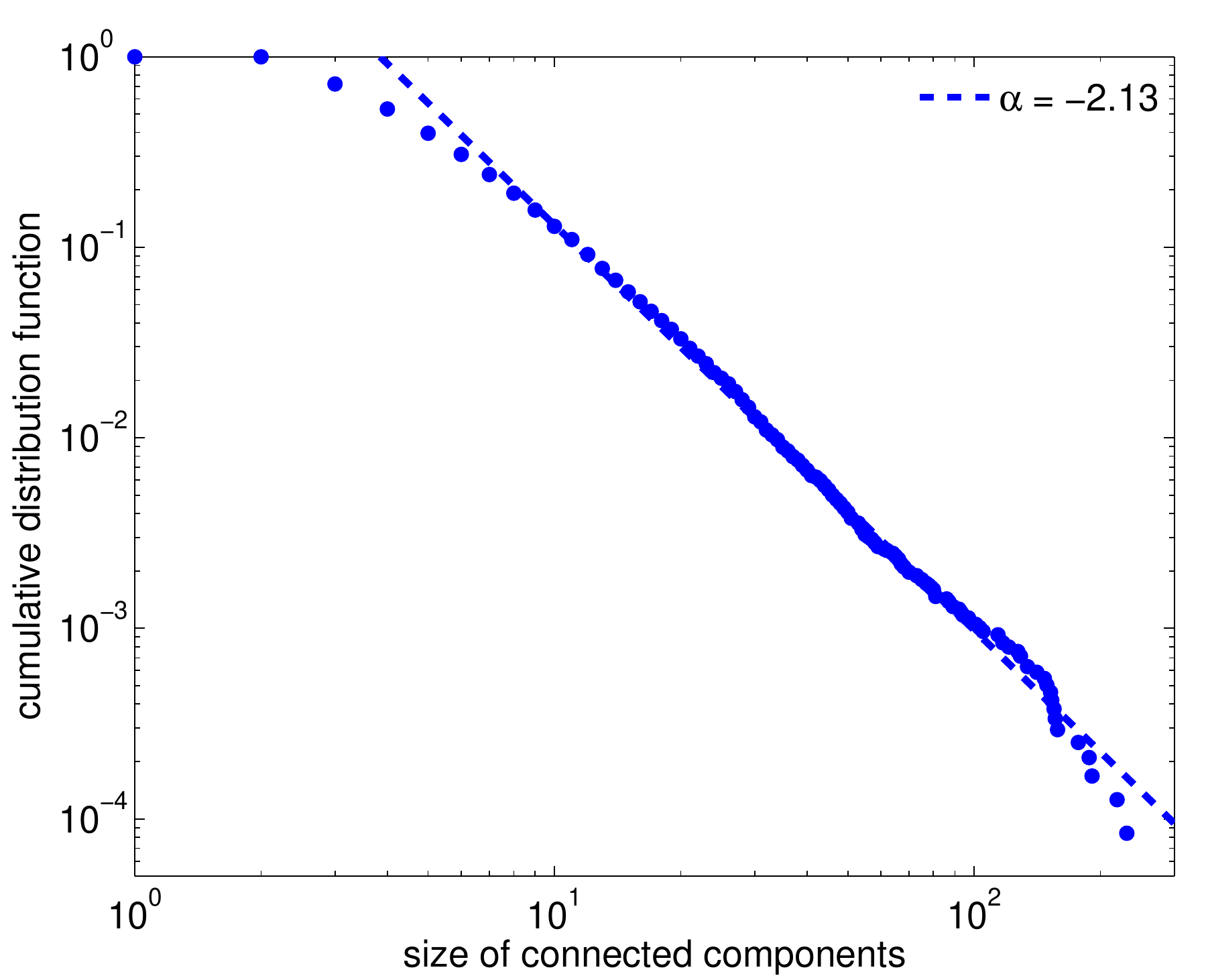}\\
   \end{center}
   \caption{ Cumulative distribution function of the size of the
     connected components. The data point representing the LCC is not shown,
     as it is three orders of magnitude larger than second largest (with
     230 nodes) and completely offset. As a comparison, a power-law with
     exponent $-3.13$ ($=\alpha-1$) is shown.} \label{sfig:cc}
\end{figure}

Ownership relations between companies create formal ties among
them. In a strongly connected component (SCC, see SI
Sec. \ref{sec:scc}), all firms reach via an ownership pathway all
other firms, thus owning each other indirectly to some extent. In
contrast, in a weakly CC firms can reach each other only if one
ignores the direction of the ownership links. This is still a
situation of interest from an economic point of view because the flow
of knowledge and information is not restricted by the direction of the
link.  The number and the size distribution of the CC provide a
measure of the fragmentation of the market.  We find that the TNC
network consists of 23825 CC. A majority of the nodes (77\%) belong to
the LCC (largest connected component) with 463006 economic actors and
889601 relations. The remaining nodes belong to CCs with sizes at
least 2000 times smaller. The second largest CC contains 230 nodes and
$90\%$ of the CC have less than 10 nodes (see Fig.  \ref{sfig:cc}).

From a geographical point of view, the LCC includes companies from 191
countries. Of these, 15491 are TNCs (about $36\%$ of all TNCs but
accounting for $94.2\%$ of the total operating revenue) from 83 different
countries. The firms that are PCs are much more numerous (399696) and are
located in only 38 countries. Finally, there are 47819 SHs from 190
countries. This means that shareholders from all around the world hold
shares in TNCs located in a more restricted number of countries, which,
in turn, further concentrates their ownership shares of PCs in an even
smaller number of countries, mainly Europe and the US.

In addition, a sector analysis of the LCC shows that the most represented
industries are the business activities sector, with 130587 companies,
followed by the services sector with 99839 companies and the
manufacturing sector with 66212 companies. On the other hand,
surprisingly, the financial intermediaries sector counts only 46632
companies. However, if we distinguish between in-going and out-going
relations, the financial intermediaries hold the largest number of shares
(341363). Instead, the manufacturing and services sectors, with
respectively 182699 and 170397 companies, have the companies with the
most shareholders.

\clearpage

\section{Bow-Tie Component Sizes}

Does a bow-tie structure and the relative size of its IN, OUT and
core result from specific economic mechanisms, or could it be explained
by a random network formation process? For correlated networks, as in
our case, there is no suitable theoretical prediction
\cite{dorogovtsev2001giant}. Heuristically, one could address the
issue by performing a random reshuffling of links. However, this would
violate economic constraints. For instance, exchanging a 10\%
ownership share in a small company with 10\% in a big one requires the
modification of the budget of the owner. In addition, the procedure is
computationally cumbersome for large data sets.

\clearpage

\section{Strongly Connected Component Analysis}
\label{sec:scc}

\begin{figure}[t]
  \begin{center}
    \begin{minipage}{0.3\textwidth}
      {\bf \textsf A} \\
\vspace{0.2in}
      \begin{center}
      \includegraphics[scale=0.8]{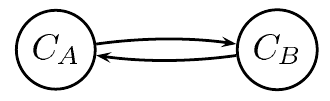}
      \end{center}
    \end{minipage}
    \begin{minipage}{0.3\textwidth}
{\bf \textsf B} 
      \begin{center}
       \includegraphics[scale=0.8]{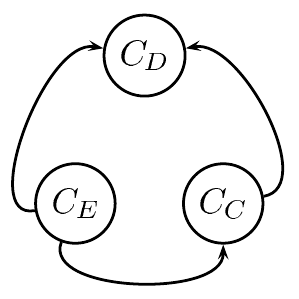}
      \end{center}
    \end{minipage}
  \begin{minipage}{0.3\textwidth}
 {\bf \textsf C}
    \begin{center}
       \includegraphics[scale=0.8]{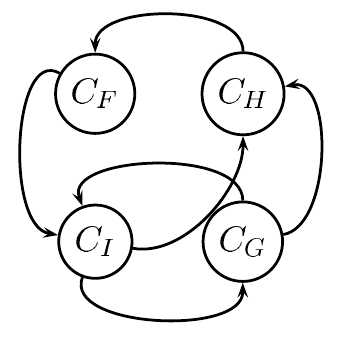}
    \end{center}
  \end{minipage}\hfill
\end{center}
\caption{ Examples of existing cross-shareholdings. ({\bf \textsf A})
  Mutual cross-shareholding. ({\bf\textsf B}) Possible cross-shareholding
  with three nodes. ({\bf \textsf C}) Cross-shareholding of higher
  degree.} \label{sfig:scceg}
\end{figure}

Cross-shareholdings, or strongly connected components (SCCs) in graph
theory, are sub-network structures where companies own each other
directly or indirectly through a chain of links (see
Fig. \ref{sfig:scceg}). Graphically speaking, this means that they form
cycles and are all reachable by every other firm in the SCC.

In economics, this kind of ownership relation has raised the attention of
different economic institutions, such as the antitrust regulators (which
have to guarantee competition in the markets), as well as that of the
companies themselves. They can set up cross-shareholdings for coping with
possible takeovers, directly sharing information, monitoring and
strategies reducing market competition.

In our sample we observe 2219 direct cross-shareholdings (4438 ownership
relations), in which 2303 companies are involved and represent 0.44\% of
all the ownership relations (see Fig. \ref{sfig:scceg}A). These direct
cross-shareholdings are divided among the different network actors as
follow:
\begin{itemize}
\item 861 between TNCs;
\item 563 between TNCs and PCs;
\item 717 between PCs;
\item 78 between SHs.
\end{itemize}
When there is a cross-shareholding involving three companies (see an
example in Fig. \ref{sfig:scceg}B), many combinations of indirect paths
are possible. In our network we observe the following ones:

\begin{itemize}
\item 829 of the type: $A \rightarrow B\rightarrow C \rightarrow A$;
\item 4.395 of the type: $A \leftrightarrow B \rightarrow C \rightarrow
  A$;
\item 8.963 of the type: $A \leftrightarrow B \leftrightarrow C
  \rightarrow A$;
\item 3.129 of the type: $A \leftrightarrow B \leftrightarrow C
  \leftrightarrow A$.
\end{itemize}

\clearpage

Next to these simple examples, we also find many SCCs with bigger
sizes. Note that smaller SCCs can be embedded in bigger ones. For
instance, in the SCC in Fig.  \ref{sfig:scceg}C there is also one
cross-shareholding between the nodes $C_I$ and $C_G$. In total there are
915 unique SCCs, of which almost all (83.7\%) are located in the
LCC. Focusing only on the LCC, there is one dominant SCC: it is
comprised of 1318 companies in 26 countries. We define the bow-tie
structure in the LCC by taking this SCC as its core (in the main text, we
only refer to this SCC). The next smallest SCC contains 286
companies. This is a group of Taiwanese firms located in the OUT of the
bow-tie. The remaining 99.7\% of SCCs in the LCC have sizes between two
and 21. The biggest SCC outside the LCC contains 19 firms.

\clearpage

\section{Network Control Concentration}

\subsection{Control of Financial Institutions}
One meaning of control in the corporate finance literature is the
frequency by which a shareholder is able to influence the firm'
strategic decision during the official voting
\cite{davis2008nfc}. Differently, in this work, by control we mean how
much economic value of companies a shareholder is able to
influence. Moreover, we did not limit our focus on the control of a
shareholder of a single firm. Instead, we look at the control each
shareholder has over its whole portfolio of directly and indirectly
owned firms. As a result, the shareholders with a high level of
control are those potentially able to impose their decision on many
high-value firms. The higher a shareholder's control is, the higher
its power to influence the final decision. In this sense, our notion of
control can be related to Weber's definition of ``power'', i.e. the
probability of an individual to be able to impose their will despite the
opposition of the others \cite{weber22som}.

In the literature on corporate control there is a debate on weather
financial institutions really exert the control associated with their
ownership shares. On the one hand, they are not supposed to seek an active
involvement in the companies' strategies. However, some works argue that
institutional investors, including banks and mutual funds, do exert
control to some extent
\cite{santos2006americansom,becht.ea05som,gillan.ea00som,davis.ea94som}. In particular, the
outcome of votes can be influenced by means of informal discussions, in
which pro-management votes are used as a bargaining chip (e.g., in
exchange of business related ``favors'' or in negotiating the extension of
credit)\footnote{For example, a mutual fund owning some percent of a
  large corporation may try to impose job cuts because of a weak economic
  situation. This can happen: (i) without voting and (ii) although the
  fund does not plan to keep these shares for many years. In this case,
  the influence of the mutual fund has a direct impact on the company and
  its employees. Furthermore, mutual funds with shares in many
  corporations may try to pursue similar strategies across their entire
  portfolio.}. On the contrary, \cite{davis.ea07som} and
\cite{davis08som} find that mutual funds, which typically hold large
blocks of shares, vote against the management (i.e., in favor of
corporate governance proposals) only 33\% of the times (in the case of
Fidelity Fund). However, they do so in more than 60\%, on average, in
other 11 cases analysed. These results are suggested to originate
mainly from a conflict of interest, where the benefits of providing
pension plan management to client corporations outweighs the possible
benefits gained from increased shareholder value. However, while some
mutual funds are reticent to exercise their power during voting mainly
in the US, an activist stance is observed for some smaller funds and
when operating outside the US \cite{davis08som}. In any case, in our
study US mutual funds represent only a small fraction of all global
financial institutions. In general, 49 mutual funds, identified by the
NACE code 6714, are among the 737 top power-holders (see main text,
Sec. Concentration of Control).

\subsection{Relation to the Rich Club Phenomenon}
The so-called rich club phenomenon \cite{colizza.ea06som,fagiolo.ea09som}
refers to the fact that in some complex networks the nodes with the
highest degree tend to be connected among each other. Being based solely
on node degree, rich club indices are not suitable for ownership
networks, in which \textit{indirect} and \textit{weighted} paths
matter. Moreover, in order to benchmark the resulting value of rich club
indices, it is usually necessary to reshuffle the links in the
network. This would be a problem in our network because it would lead to
economically unviable ownership networks. Notice, however, that the core
of the TNC network could be seen as a generalization of the rich club
phenomenon with control in the role of degree. Thus, future work should
look into this issue more in depth.

\subsection{Top Control-Holders Ranking}
This is the first time a ranking of economic actors by global control
is presented. Notice that many actors belong to the financial sector
(NACE codes starting with 65,66,67) and many of the names are
well-known global players. The interest of this ranking is not that it
exposes unsuspected powerful players. Instead, it shows that many of
the top actors belong to the core. This means that they do not carry
out their business in isolation but, on the contrary, they are tied
together in an extremely entangled web of control. This finding is
extremely important since there was no prior economic theory or
empirical evidence regarding whether and how top players are
connected. Finally, it should be noted that governments and natural
persons are only featured further down in the list.

\begin{table}[h!]
  \caption{
    Top 50 control-holders. Shareholders are ranked by network control
    (according to the threshold model, TM). Column indicate country, NACE
    industrial sector code, actor's position in the bow-tie sections, 
    cumulative network control. Notice that NACE code starting with
    65,66,67 belong to the financial sector.
  } \label{stbl:tph}
\begin{scriptsize}
  \begin{center}
    \begin{tabular}{llcccc}
      {Rank} & {Economic actor name} & { Country} & {NACE
        code} & {Network} & {Cumul. network} \\
      {     } & {                    } & {      } & {
            } & {position} & {control (TM, \%)} \\
      \hline
      1 &       BARCLAYS PLC      &	 GB      	&	6512 & 	SCC 	&	4.05	\\
      2 &       CAPITAL GROUP COMPANIES INC, THE      &	 US      	&	6713 & 	IN 	&	6.66	\\
      3 &       FMR CORP      &	 US      	&	6713  & 	IN 	&	8.94	\\
      4 &       AXA      &	 FR      	&	6712 & 	SCC 	&	11.21	\\
      5 &       STATE STREET CORPORATION      &	 US      	&	6713  &	SCC 	&	13.02	\\
      6 &       JPMORGAN CHASE \& CO.      &	 US      	&	6512 & 	SCC 	&	14.55	\\
      7 &       LEGAL \& GENERAL GROUP PLC       &	 GB      	&	6603  &	SCC 	&	16.02	\\
      8 &       VANGUARD GROUP, INC., THE      &	 US      	&	7415  &	IN 	&	17.25	\\
      9 &       UBS AG      &	 CH      	&	6512 & 	SCC 	&	18.46	\\
      10        &       MERRILL LYNCH \& CO., INC.      &	 US      	&	6712  &	SCC 	&	19.45	\\
      11        &       WELLINGTON MANAGEMENT CO. L.L.P.      &	 US      	&	6713 &	 IN 	&	20.33	\\
      12        &       DEUTSCHE BANK AG               &	DE	&	6512 &	 SCC 	&	21.17	\\
      13        &       FRANKLIN RESOURCES, INC.      &	 US      	&	6512 &	 SCC 	&	21.99	\\
      14        &       CREDIT SUISSE GROUP      &	 CH      	&	6512 &	 SCC 	&	22.81	\\
      15        &       WALTON ENTERPRISES LLC      &	 US      	&	2923 &	 T\&T 	&	23.56	\\
      16        &       BANK OF NEW YORK MELLON CORP.      &	 US	&      6512 &	 IN 	&	24.28	\\
      17        &       NATIXIS      &	 FR      	&	6512 &	 SCC 	&	24.98	\\
      18        &       GOLDMAN SACHS GROUP, INC., THE      &	 US      	&	6712 &	 SCC 	&	25.64	\\
      19        &       T. ROWE PRICE GROUP, INC.      &	 US      	&	6713 &	 SCC 	&	26.29	\\
      20        &       LEGG MASON, INC.      &	 US      	&	6712 &	 SCC 	&	26.92	\\
      21        &       MORGAN STANLEY      &	 US      	&	6712 &	 SCC 	&	27.56	\\
      22        &       MITSUBISHI UFJ FINANCIAL GROUP, INC.      &	 JP      	&	6512 &	 SCC 	&	28.16	\\
      23        &       NORTHERN TRUST CORPORATION      &	 US      	&	6512 &	 SCC 	&	28.72	\\
      24        &       SOCI\'ET\'E G\'EN\'ERALE      &	 FR      	&	6512 &	 SCC 	&	29.26	\\
      25        &       BANK OF AMERICA CORPORATION      &	 US      	&	6512 &	 SCC 	&	29.79	\\
      26        &       LLOYDS TSB GROUP PLC      &	 GB      	&	6512 &	 SCC 	&	30.30	\\
      27        &       INVESCO PLC      &	 GB      	&	6523 &	 SCC 	&	30.82	\\
      28        &       ALLIANZ SE      &	 DE      	&	7415 &	 SCC 	&	31.32	\\
      29        &       TIAA &	 US      	&	6601 &	IN 	&	32.24	\\
      30        &       OLD MUTUAL PUBLIC LIMITED COMPANY      &	 GB      	&	6601 &	 SCC 	&	32.69	\\
      31        &       AVIVA PLC      &	 GB      	&	6601 &	 SCC 	&	33.14	\\
      32        &       SCHRODERS PLC      &	 GB      	&	6712 &	 SCC 	&	33.57	\\
      33        &       DODGE \& COX      &	 US      	&	7415 &	 IN 	&	34.00	\\
      34        &       LEHMAN BROTHERS HOLDINGS, INC.      &	 US      	&	6712 &	 SCC 	&	34.43	\\
      35        &       SUN LIFE FINANCIAL, INC.      &	 CA      	&	6601 &	 SCC 	&	34.82	\\
      36        &       STANDARD LIFE PLC      &	 GB      	&	6601 &	 SCC 	&	35.2	\\
      37        &       CNCE           &	 FR      	&	6512 &	SCC 	&	35.57	\\
      38        &       NOMURA HOLDINGS, INC.      &	 JP      	&	6512 &	 SCC 	&	35.92	\\
      39        &       THE DEPOSITORY TRUST COMPANY      &	 US      	&	6512 &	 IN 	&	36.28	\\
      40        &       MASSACHUSETTS MUTUAL LIFE INSUR.      &	 US      	&	6601 &	 IN 	&	36.63	\\
      41        &       ING GROEP N.V.      &	 NL      	&	6603 &	 SCC 	&	36.96	\\
      42        &       BRANDES INVESTMENT PARTNERS, L.P.      &	 US      	&	6713 &	 IN 	&	37.29	\\
      43        &       UNICREDITO ITALIANO SPA      &	 IT      	&	6512 &	 SCC 	&	37.61	\\
      44        &       DEPOSIT INSURANCE CORPORATION OF JP      &	 JP      	&	6511 &	 IN 	&	37.93	\\
      45        &       VERENIGING AEGON      &	 NL      	&	6512 &	 IN 	&	38.25	\\
      46        &       BNP PARIBAS       &	 FR      	&	6512 &	 SCC 	&	38.56	\\
      47        &       AFFILIATED MANAGERS GROUP, INC.      &	 US      	&	6713 &	 SCC 	&	38.88	\\
      48        &       RESONA HOLDINGS, INC. &	JP	&	6512 &	 SCC 	&	39.18	\\
      49        &       CAPITAL GROUP INTERNATIONAL, INC. &	US	&     	7414 &	 IN 	&     	39.48	\\
      50        &       CHINA PETROCHEMICAL GROUP CO. &	CN	&	6511 &	 T\&T 	&	39.78	\\
      \hline
    \end{tabular}
  \end{center}
\end{scriptsize}
\end{table}

\clearpage

\section{Additional Tables}

\begin{table}[h!]
  \caption{Number of top control-holders (TCHs) located in the SCC and being members of the financial sector (FS).
    Various intersections thereof. The columns refer to the three models of network control and the TM of network value. } \label{stbl:no}
\begin{center}
  \begin{tabular}{lrrr|r} {\em } & $c^{\textrm{net}}$ {
      (LM, \#)} & $c^{\textrm{net}}$ {(TM, \#)} & $c^{\textrm{net}}$ {
      (RM, \#)} &  $v^{\textrm{net}}$ {(TM, \#)} \\ \hline
    TCH & 763 & 737 & 648 & 1791\\
    TCH$\cap$TNC & 308 & 298 & 259 & 1241\\
    TCH$\cap$TNC$\cap$SCC & 151 & 147 & 122 & 211\\
    TCH$\cap$SCC$\cap$FS & 116 & 115 & 92 & 140\\
    \hline 
  \end{tabular} 
\end{center}
\end{table}

\begin{table}[h!]
  \caption{Concentration of 80\% of network control (LM, TM,
    RM) and network value (TM). The percentages refer to the network control\/value 
    held by the TCHs according to their location in the SCC and their possible belonging to the FS,
    and various intersections thereof.} \label{stbl:nc}
\begin{center}
  \begin{tabular}{lrrr|r} {\em } & $c^{\textrm{net}}$ {
      (LM, \%)} & $c^{\textrm{net}}$ {(TM, \%)} & $c^{\textrm{net}}$ {
      (RM)} &  $v^{\textrm{net}}$ {(TM, \%)} \\ \hline
    TCH$\cap$TNC & 54.87  & 54.63  & 52.94  & 63.34 \\
    TCH$\cap$TNC$\cap$SCC & 39.54  & 38.37  & 37.29  & 30.37 \\
    TCH$\cap$SCC$\cap$FS & 36.58  & 35.37  & 34.90  & 24.36\\
    \hline 
  \end{tabular} 
\end{center}
\end{table}

\begin{table}[ht]
\caption{ Probability that a randomly chosen
  economic actor (TNC or SH) belongs to the group of top control-holders with respect to
  its position in the network structure. The first column refers to all top control-holders
  (TCHs), the second column to the first 50 TCH.} \label{stbl:prob}
\begin{center}
  \begin{tabular}{lrr}
    & { All TCH} & { First 50 TCH} \\
    \hline
    { IN} & 6.233\% & 0.273\% \\
    { SCC} & 49.831\% & 11.525\% \\
    { OUT} & 0.432\% & 0\%\\
    { T\&T} & 0.413\% & 0.002\%\\
    { OCC} & 0.016\% & 0\%\\
    \hline
  \end{tabular}
\end{center}
\end{table}


\end{document}